\documentclass[preprint,showpacs,preprintnumbers,amsmath,amssymb,prb]{revtex4-1}

\usepackage[dvipdfmx]{graphicx}
\usepackage{dcolumn}
\usepackage{bm}
\usepackage{mhchem} 
\usepackage{ascmac}
\usepackage{siunitx} 
\let\qty\SI

\usepackage{color} 
\usepackage[normalem]{ulem}
\begin{document}
%
\title{Integral-equation analysis of transient diffusion-limited currents at disk electrodes: Asymptotic expansion and compact approximation}
\author{
Kazuhiko Seki
}
\email{k-seki@aist.go.jp}
\affiliation{GZR, National Institute of Advanced Industrial Science and Technology (AIST), Onogawa 16-1 AIST West, Ibaraki, 305-8569, Japan
}
\author{Yuko Yokoyama}
\affiliation{Department of Chemical Science and Engineering, Graduate School of Engineering, Kyoto University, 
Kyotodaigaku-Katasura, Nishikyo-ku, Kyoto, 615-8510, Japan
}
\author{Masahiro Yamamoto}
\affiliation{Department of Environmental and Energy Engineering, Konan University, Okamoto 8-9-1, Hyogo, 658-8501, Japan
}
\begin{abstract}
The transient diffusion-limited current at a disk electrode following a 
change in interfacial ion concentration induced by a potential
step 
is analyzed with direct relevance to chronoamperometric measurements.
The mixed-boundary diffusion problem is formulated in the Laplace domain and reduced to a
Fredholm integral equation that directly determines the Faradaic current.
The steady-state limit recovers Saito's equation, while a systematic long-time
asymptotic expansion quantifies the approach to steady state.
A Pad\'{e} approximant yields a compact analytical expression in the
time domain that accurately describes the current over experimentally relevant
time ranges.
In contrast to existing high-accuracy numerical procedures based on hybrid
asymptotic and polynomial approximations, the present formulation provides
an explicit and compact analytical representation that facilitates interpretation
and practical implementation.
The short-time response exhibits Cottrell's equation  with edge effects
characteristic of disk electrodes.
Overall, the framework provides practical tools for analyzing transient currents,
extracting diffusion parameters, and assessing the accuracy of widely used
analytical approximations in disk-electrode chronoamperometry.
\end{abstract}

\maketitle
\newpage
\section{Introduction}
\label{sec:Intro}

Disk electrodes, particularly in the ultramicroelectrode regime, are widely used in
electrochemical measurements because they provide high current densities, rapid
attainment of steady-state conditions, and reduced ohmic and capacitive effects.
These characteristics make disk and microdisk electrodes especially suitable for
quantitative investigations of mass transport, electrode kinetics, and coupled
reaction--diffusion processes, for the detection of low-concentration species, and for
a broad range of analytical and bioelectrochemical applications.
\cite{Hapiot_08,WINLOVE_84,DENUAULT_91,O'Hare_91,Aoki_93,Arrigan_04,Evans_05,IKEUCHI_05,JIA_10,Wang_10,WANG_11,Ngamchuea_17,Sokolov_17,Noda_13,Tahara_19,BUK_19,SANDFORD_19}
These characteristics arise from enhanced radial diffusion toward the electrode, which increases the flux relative to planar geometries, and from the small electrode size, which minimizes ohmic drop and double-layer charging currents.

At the same time, ultramicroelectrodes present certain limitations. In particular, the absolute currents are small, which can impose stringent requirements on instrumentation and signal-to-noise ratios. In addition, experimental implementation may be more demanding, as the fabrication and characterization of well-defined microelectrode geometries require careful control. Variations in electrode surface condition and geometry can also affect reproducibility, particularly in quantitative measurements.

Following a potential step, the chronoamperometric current at a disk
electrode relaxes from an initial transient regime toward a finite steady-state value.
At very short times, the total current follows Cottrell's equation, exhibiting $t^{-1/2}$ time dependence, where 
$t$ denotes time,
a behavior also observed for other electrode geometries, including spherical electrodes.
At longer times, however, the disk geometry gives rise to a finite steady-state current
that reflects the mixed boundary conditions imposed by a reactive circular region
embedded in an otherwise insulating plane.

From a theoretical standpoint, diffusion to a disk electrode constitutes a mixed
boundary-value problem, with Dirichlet boundary conditions applied on the disk surface
and Neumann (insulating) boundary conditions imposed on the surrounding plane.
Such mixed boundary conditions lead to nonuniform current-density distributions
and edge-related singular behavior, particularly in the short- and intermediate-time regimes.
Analytical studies of diffusion outside conducting bodies have shown that edge effects
introduce systematic asymptotic corrections to simple one-dimensional diffusion models,
even when the total current follows Cottrell's equation.
\cite{Phillips_90_1,OLDHAM_91,Aoki_93,KANT_94}

For practical data analysis, several approximate analytical expressions have been
proposed to describe the transient diffusion-limited current at disk electrodes.
\cite{AOKI_81,SHOUP_82,AOKI_84,Mahon_05,Rajendran_99}
Among these, the Shoup--Szabo equation has become the most widely used interpolation
formula, as it provides a convenient closed-form expression that smoothly connects the
short- and long-time limits. \cite{SHOUP_82}
Despite its practical utility, experimental and numerical studies have shown that the
accuracy of this approximation depends on the time window selected for fitting, with
the largest deviations typically occurring in the intermediate-time regime most
commonly accessed in chronoamperometric experiments. \cite{IKEUCHI_05}

Exact analytical solutions for diffusion to disk electrodes can be formulated in terms
of special functions, such as spheroidal wave functions; however, their mathematical
complexity and the need for numerical evaluation of infinite series limit their direct applicability in routine data analysis.
\cite{Rajendran_99,BIENIASZ_16,BIENIASZ_18}
Although a formal exact solution exists in the Laplace domain, 
its practical evaluation requires numerical treatment of special functions and infinite series.
At the same time, highly accurate and computationally efficient numerical procedures have been developed.
In particular, the approach reported in Ref.~\onlinecite{BIENIASZ_18} enables the evaluation of the transient current
with near machine precision at very low computational cost by combining asymptotic expansions
with polynomial approximations based on reference data.
While such methods are extremely effective for numerical evaluation, they are inherently algorithmic
and do not yield compact closed-form expressions that can be readily used for analytical manipulation,
parameter estimation, or theoretical interpretation.
This situation motivates the development of theoretical frameworks that treat the mixed
boundary conditions rigorously while remaining amenable to numerical evaluation and
systematic approximation.
In particular, there remains a need for simple analytical representations that provide
transparent functional forms and can be readily incorporated into theoretical models
and data analysis procedures.
Such approaches are particularly valuable in modern micro- and nanoelectrode
applications, where accurate interpretation of transient currents is essential for
establishing mass-transport control.
\cite{Yu_17,Adachi_20,Conceicao_24}
The results obtained in the present work are consistent with high-accuracy numerical evaluations such as those reported in Ref.~\onlinecite{BIENIASZ_18}, which serve as a benchmark for assessing the present analytical approximations.
The present approach is therefore complementary to high-accuracy numerical schemes, providing analytically tractable expressions with clear physical interpretation.

In the present work, we analyze the relaxation of the diffusion-limited current at a disk electrode of radius 
$a$ following a change in interfacial ion concentration induced by a potential step at $t=0$. 
By formulating the diffusion problem in the Laplace domain, the governing equations are
reduced to a modified Helmholtz equation and subsequently to a Fredholm integral
equation of the second kind for an auxiliary function that directly determines the
total Faradaic current.
This formulation enables accurate numerical evaluation of the transient current and
the systematic derivation of long-time asymptotic expansions.
In addition, a Pad\'{e} resummation in the Laplace domain is introduced to obtain a
compact closed-form approximation that accurately describes the current over
intermediate and long times.

It is important to distinguish between the formally exact short-time limit and
the transient regimes that are experimentally accessible in
chronoamperometric measurements at disk electrodes.
An idealized instantaneous potential step leads, in the
limit $t \to 0^{+}$, to a Cottrell-type current proportional to the electrode
area, reflecting locally planar diffusion.
In practice, however, this regime is often obscured by finite instrumental
response times, double-layer charging, and other non-diffusive effects.
\cite{Aoki_93,Amatore_07}
Consequently, the experimentally relevant transient response is typically
governed by the approach to the finite steady-state current of a disk
electrode, 
where the leading correction again exhibits a Cottrell-type decay with $t^{-1/2}$ time dependence and a distinct amplitude.
This regime underlies widely used analytical approximations for disk
electrodes and motivates a unified theoretical description that consistently
connects the short-time, long-time, and steady-state limits.

Overall, the approach presented here provides a unified and mathematically
transparent description of steady-state, transient, and short-time behavior
at disk electrodes, clarifies the role of edge effects arising from mixed
boundary conditions, and offers practical analytical and numerical tools
for interpreting chronoamperometric data beyond existing interpolation formulas.

\section{Formulation}
\label{sec:Formulation}

\begin{figure}[h]
\begin{center}
\includegraphics[width=\textwidth]{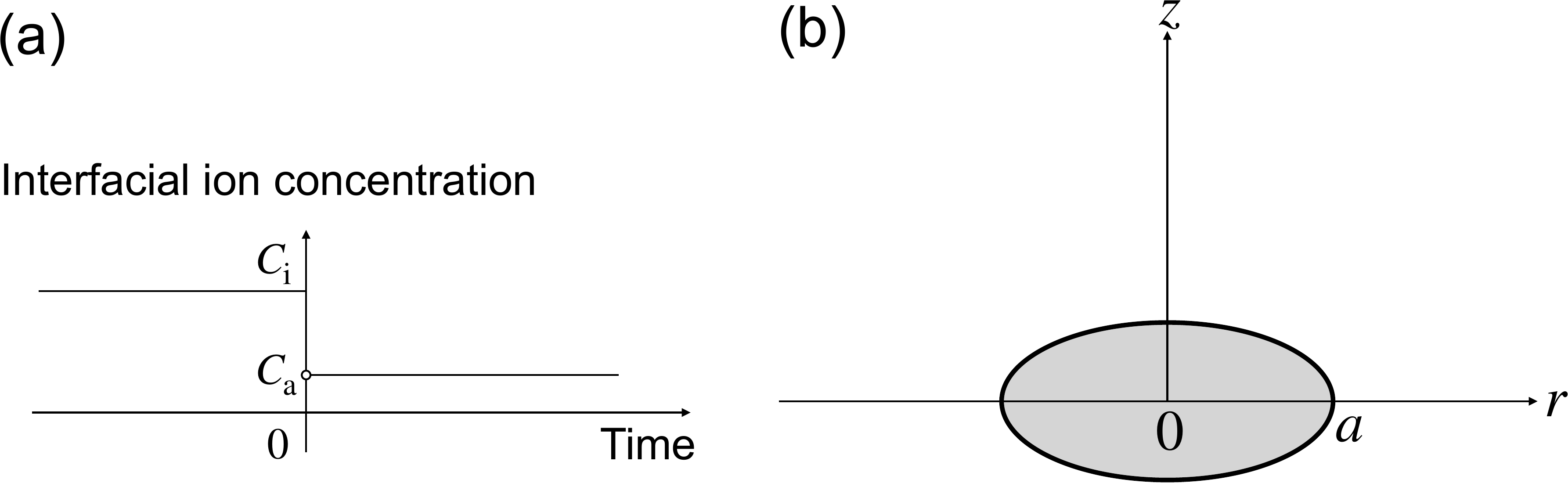}
\end{center}
\caption{
(a) Initial uniform concentration $C_{\rm i}$. 
At $t = 0^{+}$, the interfacial ion concentration is changed to $C_{\rm a}$ by applying a potential step and is maintained at this value thereafter. 
We define $\Delta C(r,z,t) = C(r,z,t) - C_{\rm i}$ and $C_0 = C_{\rm a} - C_{\rm i}$. 
(b) Cylindrical coordinate system $(r,z)$ used for diffusion modeling and flux/current calculations at a disk electrode of radius $a$. 
If $I(t) < 0$, the current flows in the negative $z$-direction.
}
\label{fig:disk}
\end{figure}

We consider a system that is initially under non-reactive conditions for $t < 0$, with a uniform concentration $C_{\rm i}$ (Fig.~\ref{fig:disk}(a)). 
At $t = 0^{+}$, the interfacial ion concentration is changed to $C_{\rm a}$ by applying a potential step. 
We study the subsequent current relaxation when the concentration $C(r,0,t)$ within the circular region of radius $a$ at $z=0$ is maintained at $C_{\rm a}$ for $t>0$.

Aside from the initial charging of the electric double layer, the current $I(t)$ is diffusion-controlled. 
We impose a Dirichlet boundary condition at the electrode surface, $C(r,0,t)=C_{\rm a}$, 
which corresponds to sufficiently fast electron-transfer kinetics (Nernstian equilibrium) at the interface. 
Under this condition, the surface concentration is fixed by thermodynamic equilibrium when $C_{\rm a} > 0$, 
and transport remains diffusion-limited. 
In addition, no homogeneous chemical reaction between reactant and product species in the solution is assumed, 
except in Sec.~9, where the relation to a diffusional EC$'$ mechanism is discussed. 
Accordingly, a single diffusion equation provides an adequate description. 
In the limiting case of a fast irreversible reaction, $C_{\rm a} \approx 0$. 
The diffusion-controlled current $I(t)$ relaxes to the steady-state value $I_{\infty} $. 
We examine the relaxation function $I(t)/I_{\infty} $, which is independent of $C_{\rm i}$ and $C_{\rm a}$ owing to the linearity of the diffusion equation. 
For convenience, we define $\Delta C(r,z,t) = C(r,z,t) - C_{\rm i}$.

We analyze the transport of charge carriers to a circular reactive region (disk) of radius $a$ located in the plane $z=0$. 
The plane $z=0$ is insulating except for the disk; that is, there is no normal flux across $z=0$ outside the reactive area. 
The origin is placed at the disk center, and $r$ denotes the radial coordinate in the plane $z=0$ (Fig.~\ref{fig:disk}(b)). 
The concentration field $C(r,z,t)$ is considered in the upper half-space, $z \ge 0$. 
Assuming rotational symmetry about the $z$-axis, $C(r,z,t)$ satisfies Fick's second law,
\begin{align}
\frac{\partial C(r,z,t)}{\partial t} = D \nabla^2 C(r,z,t),
\label{eq:B1t}
\end{align}
where the Laplacian under rotational symmetry is given by
\begin{align}
\nabla^2 C(r,z,t) = \Delta_2 C(r,z,t) + \frac{\partial^2 C(r,z,t)}{\partial z^2},
\label{eq:B1_1}
\end{align}
with
\begin{align}
\Delta_2 C(r,z,t) = \frac{1}{r}\frac{\partial}{\partial r}
\left(
r \frac{\partial C(r,z,t)}{\partial r}
\right).
\label{eq:B3}
\end{align}

We denote the Laplace transform of $f(t)$ by $\hat{f}(s)$. 
As shown in Appendix~A, the concentration in the upper half-space under the circular boundary conditions at $z=0$ can be expressed as \cite{Duffy_08}
\begin{align}
\Delta \hat{C}(r,z,s) =
\int_0^\infty dk \, A(k,s) J_0(kr)
\exp\!\left(-z\sqrt{k^2 + s/D}\right),
\label{eq:T1}
\end{align}
where 
$k$ is the separation constant arising from the method of
separation of variables and serves as the radial transform variable in the integral representation of the solution, $A(k,s)$ is determined by the boundary
conditions, and  
$J_n(z)$ denotes the Bessel function of the first kind \cite{NIST}. 
By definition, $\Delta C(r,z,t) \to 0$ as $r \to \infty$ and/or $z \to \infty$; thus, $C(r,z,t) \to C_{\rm i}$ in these limits, where the influence of the circular electrode becomes negligible.

\subsection*{Faradaic current}
To avoid confusion between diffusive flux and electrical current, we use $\vec{{\cal J}}$ for the diffusive molar flux density (units of amount\,area$^{-1}$\,time$^{-1}$) and $I$ for the total Faradaic current (units of charge\,time$^{-1}$).
The Faradaic conversion is performed by the factor $nF$, where 
$F$ indicates the Faraday constant, $n$ is the number of electrons transferred. 

We define the diffusive flux density vector,
\begin{align}
\hat{\vec{{\cal J}}}=-D \mbox{grad} \Delta \hat{C}(r,z,s) .
\label{eq:B5}
\end{align}
The flux in the $z$-direction can be expressed as
\begin{align}
\hat{{\cal J}}_z (r,z,s) = -D\frac{\partial}{\partial z} \Delta \hat{C}(r,z,s) .
\label{eq:B6}
\end{align}
By substituting Eq. (\ref{eq:T1}) into Eq. (\ref{eq:B6}), we obtain
\begin{align}
\lim_{z\rightarrow +0} \hat{{\cal J}}_z (r,z,s)  &=\left.  -D\frac{\partial}{\partial z} \int_0^\infty dkA(k,s) J_0(kr) \exp\left(
-z\sqrt{k^2+s /D} 
\right) \right|_{z=+0}
\label{eq:B8}\\
&=D\int_0^\infty dk\sqrt{k^2+s /D} \, A(k,s) J_0(kr) . 
\label{eq:B6_1}
\end{align}

The total Faradaic current through the circular region at $z=0$ can be calculated from
\begin{align}
\hat{I} (s) &= 2\pi nF\int_0^a r dr \,\hat{{\cal J}}_z (r,z=+0,s)
\label{eq:B7}\\
&=2\pi DnF\int_0^a r dr \int_0^\infty dk\sqrt{k^2+s /D} \, A(k,s) J_0(kr) 
\label{eq:B9_0}\\
&=2\pi D a nF\int_0^\infty dk\frac{\sqrt{k^2+s /D}}{k} \, A(k,s) J_1(k a) ,
\label{eq:B10}
\end{align}
where we have used (10.22.1 of Ref. \onlinecite{NIST})
\begin{align}
\int_0^a dr \, r J_0(kr)=\frac{a}{k} J_1 (ka). 
\label{eq:B9}
\end{align}
If $I(t)$ is negative, the current flows in the negative direction of the $z$-axis in the upper half-space, $z \ge 0$.

Introducing $C_0 = C_{\rm a} - C_{\rm i}$ and $\Delta C(r,z,t) = C(r,z,t) - C_{\rm i}$, 
the boundary condition inside the disk is expressed as $\Delta C(r,0,t) = C_0$ for $0 \leq r < a$. 
Outside the disk, the plane is insulating; that is, $\partial \Delta C(r,z,t)/\partial z = 0$ at $z = 0$ for $r > a$.
In the Laplace domain, $\Delta \hat{C}(r,z,s)$ satisfies the boundary conditions
\begin{align}
\Delta \hat{C}(r,0,s) &= \frac{C_0}{s} \quad \text{for } 0 \leq r < a ,
\label{eq:T15}\\
\lim_{z \to +0} \hat{{\cal J}}_z (r,z,s) &= 0 \quad \text{for } a < r < \infty .
\label{eq:T16}
\end{align}Using Eqs. (\ref{eq:T1}) and (\ref{eq:B6_1}), 
we note that $A(k,s)$ satisfies the dual integral equations implied by these boundary conditions, \cite{Duffy_08}
\begin{align}
\int_0^\infty dk A(k,s) J_0(kr) &=C_0/s \mbox{ for } 0 \leq r < a ,
\label{eq:T2}\\
\int_0^\infty dk \sqrt{k^2+s /D}\, A(k,s) J_0(kr) &=0\mbox{ for } a < r < \infty .
\label{eq:T3}
\end{align}

To reduce these equations to a form for which a standard solution method is available, we introduce $\hat{g}(k,s)$ satisfying \cite{Duffy_08}
\begin{align}
\hat{g}(k,s) =\sqrt{k^2+s /D}A(k,s)/C_0 ,
\label{eq:T4}
\end{align}
and rewrite Eqs. (\ref{eq:T2}) and (\ref{eq:T3}) as \cite{Duffy_08}
\begin{align}
\int_0^\infty d k \frac{\hat{g}(k,s)}{\sqrt{k^2+s /D}} \, J_0(k r) &=1/s \mbox{ for } 0 \leq r < a ,
\label{eq:T5}\\
\int_0^\infty d k \, \hat{g}(k,s) J_0(k r ) &=0 \mbox{ for } a < r < \infty .
\label{eq:T6}
\end{align}

A method for solving the dual integral equations [Eqs.~(\ref{eq:T5}) and (\ref{eq:T6})] has been established \cite{COOKE_56,Sneddon_66,Duffy_08,Agra_04,Yamamoto_24}.
Although Eqs.~(\ref{eq:T5}) and (\ref{eq:T6}) can be interpreted as inverse Hankel transforms, a more effective method was proposed by Cooke.
According to Cooke (1956) (see p.~213 of Ref.~\onlinecite{Duffy_08}) \cite{COOKE_56}, we express $\hat{g}(k,s)$ in terms of $\hat{h}(r,s)$ as
\begin{align}
\hat{g}(k,s) &= \frac{2}{\pi} k \int_0^a \hat{h}(r,s) \cos(kr)\,dr.
\label{eq:F16}
\end{align}
We show below that $\hat{h}(r,t)$ gives the $z$-component of the flux at the boundary, and hence determines the Faradaic current.

The total current flowing through the circular region of radius $a$ at $z=0$, 
Eq. (\ref{eq:B10}), can be expressed as
\begin{align}
\hat{I} (s)&=2\pi D a nF C_0 \int_0^\infty d k \hat{g}(k,s) \frac{1}{k} J_1 (ka) . 
\label{eqF2}
\end{align}
Using $\hat{h}(r,s)$, $\hat{I} (s)$ can be expressed as
\begin{align}
\hat{I} (s)&=4D a nF C_0 \int_0^a dr \hat{h}(r,s)\int_0^\infty d k J_1 (ka) \cos(k r)
\label{eqF3_0}\\
&=4D nF C_0 \int_0^a dr \hat{h}(r,s) ,
\label{eqF3}
\end{align}
where we have used 
$\int_0^\infty d k   J_1 (ka)\cos (kr)=1/a$ 
for $0<r<a$. (6.693.1 of Ref. \onlinecite{gradshteyn_96})
Equation (\ref{eqF3}) indicates that $h(r,t)$ determines the radial
distribution of the flux contribution to the total Faradaic current
at position $r$ on the boundary $z=0$.

\section{Steady State}
\label{sec:Steady state}

Results in the steady state 
($t\rightarrow \infty$) can be obtained by taking the limit $s\rightarrow 0$ and applying the inverse Laplace transformation;  
we note that the inverse Laplace transform of $1/s$ is $1$. 

In the limit $s\rightarrow 0$, we define
\begin{align}
\hat{g}_0(k) =\lim_{s\rightarrow 0} \sqrt{k^2+s /D}A(k,s)/C_0=k A(k,0)/C_0 . 
\label{eq:SS1}
\end{align}

By taking the limit $s\rightarrow 0$, and applying the inverse Laplace transformation, Eqs. (\ref{eq:T5}) and (\ref{eq:T6}) become
\begin{align}
\int_0^\infty d k \frac{g_0 (k) }{k} \, J_0(k r) &=1 \mbox{ for } 0 \leq r < a ,
\label{eq:T5_1}\\
\int_0^\infty d k \, g_0(k)  J_0(k r ) &=0 \mbox{ for } a < r < \infty . 
\label{eq:T6_1}
\end{align}

Using the formula 
(6.693.6 and 6.671.7 of Ref. \onlinecite{gradshteyn_96})
\begin{align}
\int_0^\infty \frac{dk}{k} \, \sin(k a) J_0(k r )&=
\pi/2  &\mbox{ for }0 \leq r < a,
\label{eq:SS2}\\
\int_0^\infty dk \sin(k a) J_0(k r )&=0                  &\mbox{ for }a < r < \infty ,
\label{eq:SS3}
\end{align}
we obtain
\begin{align}
g_0(k)=\frac{2}{\pi} \sin (ka) .
\label{eq:SS4}
\end{align}
The analytical expression for the total current flowing through the circular region of radius $a$ 
[the inverse Laplace transform of Eq. (\ref{eqF2})] is then
\begin{align}
I_{\infty} &=2\pi D a nF C_0\int_0^\infty d k g_0 (k) \frac{1}{k} J_1 (ka) 
=4D a nF C_0 ,
\label{eq:SS7}
\end{align}
where we have used 
$\int_0^\infty d k  \sin (ka) J_1 (ka)/k=1$. (6.693.1 of Ref. \onlinecite{gradshteyn_96}) 
Equation~(\ref{eq:SS7}) corresponds to Saito's equation. \cite{SAITO_68}
When formulated in terms of electrical or ionic conductivity (or, equivalently,
resistivity), this expression is recognized as  the Holm constriction (spreading) resistance or,
alternatively,  the Maxwell--Hall access resistance. 
\cite{Holm,Newman_66,maxwell_54,Hall_75,Britz_16} 
(For the rate of heat flow, see Sec. 8.2 of Ref. \onlinecite{Carslaw_59})
This expression has also been extended to describe anisotropic layers in the context of
electrical conductivity. \cite{Koren_14,Koren_14_1,Seki_Kubo_20,Seki_20,Seki_Kubo_21,Seki_21}
Here, we apply the method of solving mixed boundary value problems, originally developed to
analyze electrical contacts in anisotropic layers, to the relaxation of diffusive
currents toward the steady-state current described by Saito's equation.

\section{Time dependent solution: derivation}
\label{sec:Derivation}

We study the transient relaxation of the current to the steady state given by Eq. (\ref{eq:SS7}), {\it i.e.}, Saito's equation. 
The results of this section are a special case of the general solution given in the Appendix C. 
Nevertheless, we provide the derivation for completeness. 
We also reformulate the problem using dimensionless variables in Appendix~C.

We first substitute Eq. (\ref{eq:F16}) into Eq. (\ref{eq:T6}) by rewriting Eq. (\ref{eq:F16}) as
$\hat{g}(k,s)=(2/\pi) k\int_0^a dy \hat{h}(y,s) \cos(k y)$, 
\begin{align}
\int_0^a dy \hat{h}(y,s) \int_0^\infty d k \, \frac{2}{\pi} k\cos(ky) J_0(k r ) &=0 \mbox{ for } a < r < \infty .
\label{eq:T6F16}
\end{align}
The boundary condition given by Eq. (\ref{eq:T6F16}) is satisfied irrespective of the functional form of $\hat{h}(r,s)$, since 
$\hat{h}(r,s)$ is independent of $k$ and
\begin{align}
\int_0^\infty dk \, k \cos(k y) J_0(k r )=0  \mbox{ for } y < r < \infty ,
\label{eq:T11}
\end{align}
which follows from (10.22.59 of Ref. \onlinecite{NIST}) 
\begin{align}
\int_0^\infty dk \, \sin(k y) J_0(k r )=0  \mbox{ for } y < r < \infty ; 
\label{eq:T12}
\end{align}
we then have
\begin{align}
\frac{d}{dy}\int_0^\infty dk \, \sin(k y) J_0(k r )=0  \mbox{ for } y< r < \infty .
\label{eq:T13}
\end{align}
Noting that $y< r$ in Eq. (\ref{eq:T6F16}), we obtain Eq. (\ref{eq:T11}). 
We have thus shown that the boundary condition in Eq. (\ref{eq:T6}) is satisfied irrespective of the functional form of $\hat{h}(r,s)$.

Next, we substitute Eq. (\ref{eq:F16}) into Eq. (\ref{eq:T5}), 
\begin{align}
\frac{2}{\pi} \int_0^a dy \hat{h}(y,s) \int_0^\infty d k \cos(ky)\frac{k}{\sqrt{k^2+s /D}} \, J_0(k r) &=1/s \mbox{ for } 0 \leq r < a ,
\label{eq:T5F16}
\end{align}
The function $\hat{h}(r,s)$ satisfying the other boundary condition given by Eq. (\ref{eq:T5}) must satisfy 
the Fredholm integral equation of the first kind for $\hat{h}(y,s)$, given by 
Eq. (\ref{eq:T5F16}). 

As shown in Appendix B, 
Eq. (\ref{eq:T5F16}) can be rewritten as 
a Fredholm integral equation of the second kind for $\hat{h}(y,s)$ 
\cite{Duffy_08}
\begin{align}
\hat{h}(r,s)= \frac{1}{s}+
\int_0^a dy\, \hat{h}(y,s)\hat{K}(r,y,s)
\quad \text{for } 0 \leq r < a,
\label{eq:F17}
\end{align}
where the kernel is defined as
\begin{align}
\hat{K}(r,y,s)=-\frac{2}{\pi}
\int_0^\infty dk 
\left[\frac{k}{\sqrt{k^2+s/D}}-1 
\right]
\cos(k y)\cos(k r)
\quad \text{for } 0 \leq r < a.
\label{eq:asymex1}
\end{align}
Equations~(\ref{eq:F16}) and (\ref{eq:F17}), together with Eq.~(\ref{eq:asymex1}),
can be derived by applying the general method described in Appendix~C, {\it i.e.},
Eqs.~(\ref{eq:gm5}) and (\ref{eq:gm6}).

Since the total current is obtained by spatial integration of $h(r,t)$ over the disk
area,
$I(t)=4 D nF C_0 \int_0^a dr\, h(r,t)$ [Eq.~(\ref{eqF3})],
$h(r,t)$ is proportional to the local contribution to the flux-weighted current density normal to the disk surface at radial
position $r$ and time $t$.
The first term on the right-hand side of Eq.~(\ref{eq:F17}) corresponds to the steady-state
current, while the integral term describes the temporal evolution toward this steady
state.
Specifically, the integral term can be interpreted as a memory effect; the local flux
at position $r$ and time $t$ depends on the distribution of fluxes at earlier times
$t' < t$ over the entire disk.

This nonlocal temporal coupling is mediated by the kernel $K(r,y,t-t')$, which quantifies
how a current density at position $y$ and time $t'$ contributes to the response at position
$r$ at a later time $t$.
In the Laplace domain, this memory kernel is represented by $\hat{K}(r,y,s)$.
Thus, Eq.~(\ref{eq:F17}) explicitly expresses the relaxation process as a
spatio-temporal convolution, reflecting the diffusive coupling between different regions
of the electrode surface.

By substituting $\hat{h}(r,s)$ determined by solving Eq. (\ref{eq:F17}) with Eq. (\ref{eq:asymex1}) 
into Eq. (\ref{eqF3}), we obtain the total current flowing through the circular region of radius $a$ at $z=0$ in the Laplace domain. 
To calculate $\hat{I} (s)$, we need $\hat{h}(r,s)$ in the integrand of $\int_0^a dr \cdots$, 
which indicates that $\hat{h}(r,s)$ for $0 \leq r < a$ is required. 
Therefore, we drop the condition $0 \leq r < a$ for $\hat{h}(r,s)$ below. 
Equation (\ref{eq:F17}) involves double integrations. 
However, Eq. (\ref{eq:asymex1})  can be expressed by analytical function, \cite{Mathematica}
\begin{multline}
\hat{K} (r,y,s)
=
\frac{1}{2} \sqrt{\frac{s}{D}}\left[\frac{4}{\pi}+ 
\left(\bm{L}_1 \left(\sqrt{\frac{s}{D}} (r+y)
\right) -I_1 \left(\sqrt{\frac{s}{D}} (r+y)
\right)
\right) 
\right.\\
\left.
+\left(\bm{L}_1 \left(\sqrt{\frac{s}{D}} (r-y)
\right) -I_1 \left(\sqrt{\frac{s}{D}} |r-y|
\right)\right)
\right],
\label{eq:F1}
\end{multline}
where $\bm{L}_n(z)$ indicates the modified Struve function, 
and $I_n(z)$ indicates the modified Bessel function of the first kind. \cite{NIST}  
We note that $\bm{L}_1(z)$ is an even function. \cite{NIST}

Equations~(\ref{eq:F16}) and (\ref{eq:F17}), together with Eq.~(\ref{eq:asymex1}), have already been reported in Ref.~\onlinecite{Duffy_08}.
By using Eq.~(\ref{eq:F1}), Eq.~(\ref{eq:F17}) can be discretized more straightforwardly than performing the numerical integration required in Eq.~(\ref{eq:asymex1}).
In the following, we explicitly solve Eqs.~(\ref{eq:F16}) and (\ref{eq:F17}) using Eq.~(\ref{eq:F1}).
The numerical solution of Eq. (\ref{eq:F17}) is substituted into 
Eq. (\ref{eqF3}); we obtain the Laplace transform of the total current flowing through the circular region of radius $a$ at $z=0$. 
The time dependence is obtained by applying a numerical inverse Laplace transform to Eq. (\ref{eqF3})
using the Stehfest algorithm. \cite{Stehfest1970_47,Stehfest1970_624}

In the following, we distinguish between the steady-state solution and the time-dependent asymptotic solution. 
From a formal mathematical perspective, the steady-state solution corresponds to the leading-order term in the large-time asymptotic expansion. 
However, we use the term ``steady-state solution'' to refer specifically to the strictly time-independent limit, 
whereas the ``asymptotic solution'' describes the time-dependent behavior approaching this limit.

\section{Steady state: revisited}
\label{sec:Revisited}

We define $\hat{h}_0(r)=\lim_{s\rightarrow 0} \hat{h}(r,s)$, and $\hat{g}_0(r)=\lim_{s\rightarrow 0} \hat{g}(r,s)$. 
Using Eq. (\ref{eq:F17}), we find $\hat{h}_0(r)=1/s$ and $h_0(r)=1$ after applying the inverse Laplace transformation.
Substituting $\hat{h}_0(r)=1/s$ into Eqs. (\ref{eq:F16})
and (\ref{eqF3}), we obtain $\hat{g}_0(r)=(2/\pi) \sin (ka)/s$, 
$g_0(k)=(2/\pi) \sin (ka)$ [Eq. (\ref{eq:SS4})], and $I_{\infty}=4DanFC_0$. 
Therefore, Saito's equation [Eq. (\ref{eq:SS7})] is derived. 

\section{Long-time asymptotic solution}

We consider the exact integral equation, Eq. (\ref{eq:F17}), 
where $\hat{K} (r,y,s)$ is the kernel defined in
Eq.~(\ref{eq:asymex1}).
We first expand $\hat{K} (r,s)$ using Eq. (\ref{eq:F1}) by regarding $s a^2/D$ as a small parameter and assuming $r/a \leq 1$ and $y/a \leq 1$; 
the kernel $\hat{K} (r,y,s)$ admits the series representation
\begin{align}
\hat{K}(r,y,s)=
\sum_{n=1}^\infty K^{(n)} (r,y)\left( \frac{s}{D} \right)^{n/2},
\label{eq:Ga2}
\end{align}
with coefficient functions $K^{(n)} (r,y)$ given by
\begin{multline}
K^{(n)}  (r,y) =\left[
\frac{1-(-1)^n}{2} a_{(n-3)/2} -
\frac{1+(-1)^n}{2} b_{n/2}
\right] \times \\
\left[\left(r+y\right)^{n-1}
+\frac{1-(-1)^n}{2} (r-y)^{n-1}+
\frac{1+(-1)^n}{2} |r-y|^{n-1}
\right] ,
\label{eq:Ga4}
\end{multline}
where we defined, 
\begin{align}
a_n &=\frac{2^{2n+1}n! (n+1)!}{\pi (2n+1)! (2n+3)!} ,
\label{eq:ac1}
\\
b_n &=\frac{n}{2^{2n} \left( n!\right)^2} , 
\label{eq:bc1}
\end{align}
and $a_{-1}=1/\pi$.
The factors $[1-(-1)^n]/2$ and $[1+(-1)^n]/2$ select the even- and odd-$n$
contributions, respectively.

Corresponding to the expansion in powers of $(s/D)^{1/2}$ in Eq. (\ref{eq:Ga2}), the solution $\hat{h}(r,s)$ may admit the following asymptotic expansion, 
\begin{align}
\hat{h}(r,s)=\sum_{\ell=0}^\infty h^{(\ell)} (r)\frac{1}{s}\left(
\frac{sa^2}{D}
\right)^{\ell/2} .
\label{eq:Ga5}
\end{align}
The $\ell$th-order iterated function $h^{(\ell)} (r)$ is obtained by substituting
Eqs.~(\ref{eq:Ga2}) and (\ref{eq:Ga5}) into the integral equation, yielding
\begin{align}
h^{(\ell)} (r) &= \sum_{n=1}^\ell \int_0^a dy\, h^{(\ell-n)} (y)\,
K^{(n)} (r,y)/a^n ,
\qquad \ell\geq 1,
\label{eq:Ga6}
\end{align}
with the initial condition $h^{(0)} (r)=1$ and $K^{(n)}  (r,y)$ given by Eq. (\ref{eq:Ga4}).
In Appendix~D, this recursive relation is implemented efficiently using
pre-stored moment integrals.

Using the iterated functions obtained from Eq.~(\ref{eq:Ga6}), 
$\hat{h}(r,s)$ is derived from Eq.~(\ref{eq:Ga5}). 
Applying the inverse Laplace transform via the Tauberian theorem, \cite{Feller_71} 
while neglecting the even-power terms in $sa^2/D$ that do not contribute to the asymptotic expansion in the time domain, 
yields
\begin{align}
h(r,t) = 1 +
\sum_{m=0}^{\infty} \frac{h^{(1+2m)}(r)}{\Gamma(1/2 - m)}
\left( \frac{a^2}{Dt} \right)^{1/2 + m}.
\label{eq:Ga10}
\end{align}

The transient current follows from Eq.~(\ref{eqF3}) as
\begin{align}
I (t)&=4D nF C_0 \int_0^a dr\, h(r,t)
\nonumber\\
&=4D a nF C_0 
\left[1+\sum_{m=0}^\infty
\frac{\int_0^a dr\, h^{(1+2m)} (r)}{a \Gamma(1/2-m)} \left( 
\frac{a^2}{Dt}
\right)^{1/2+m} \right],
\label{eq:Ga11}
\end{align}
while the corresponding expression in the Laplace domain is
\begin{align}
\hat{I} (s)&=\frac{4D a nF C_0}{s} 
\left[1+\sum_{m=0}^\infty \frac{1}{a}
\int_0^a dr\, h^{(1+2m)} (r) \left( 
\frac{sa^2}{D}
\right)^{1/2+m} \right].
\label{eq:Ga11s}
\end{align}
For convenience, we introduce, 
\begin{align}
\frac{I^{(1+2\ell)}}{I_{\infty}}=\frac{1}{a}
\int_0^a dr\, h^{(1+2m)} (r) , 
\label{eq:j_ell}
\end{align}
where $I_{\infty}=4DanFC_0$. 

As shown in Appendix D, Eqs.~(\ref{eq:Ga11}) and (\ref{eq:Ga11s}) reproduce the asymptotic results 
obtained by evaluating the exact expression
for the current expanded in terms of radial spheroidal wave functions of the third
kind, \cite{BIENIASZ_18}
and provide a unified
framework for generating higher-order correction terms.
In  Appendix~D, 
higher-order correction terms obtained using this approach are also presented. 

By applying the systematic asymptotic expansion up to the 9th iteration from the steady state in the long-time limit, 
we obtain
\begin{multline}
\frac{I_{\rm ip}(t)}{ I_{\infty} } =  1+ \frac{2a}{\pi^{3/2} \left( Dt \right)^{1/2}}
+\left(\frac{1}{9}-\frac{1}{\pi^2} \right) \frac{4a^3}{\left(\pi D t\right)^{3/2}} +
\left(\frac{71}{450 }-\frac{4}{\pi ^{2}}+\frac{24}{\pi ^{4}}\right) \frac{a^5}{\pi^{3/2}(Dt)^{5/2}}+
\\
\left(\frac{71}{2205}-\frac{242}{135 \pi ^{2}}+\frac{80}{3 \pi ^{4}}-\frac{120}{\pi ^{6}}\right)
\frac{2a^7}{\pi^{3/2}(Dt)^{7/2}}+\\
\left(\frac{9523}{340200}-\frac{5396}{1701 \pi ^{2}}+\frac{266}{3 \pi ^{4}}-\frac{2800}{3 \pi ^{6}}+\frac{3360}{\pi ^{8}}
\right)\frac{a^9}{\pi^{3/2}(Dt)^{9/2}} .
\label{eq:asymanal1}
\end{multline}
Here $I_{\infty} =4D a nF C_0$ [Eq. (\ref{eq:SS7})] and  $I_{\rm ip}(t)$ indicates the results of inverse power series expansion. 
The first three terms on the right-hand side of Eq. (\ref{eq:asymanal1}) coincide with the known results derived by a different method; \cite{SHOUP_82}
$1+2a/(\pi^{3/2} (Dt)^{1/2})$ was derived by Shoup and Szabo 
by correcting an error in the original derivation by Aoki and Osteryoung using the Wiener--Hopf method,  
\cite{AOKI_81,SHOUP_82,AOKI_84}
and the third term $\left(1/9-1/\pi^2 \right) 4a^3/\left(\pi D t\right)^{3/2} $ was also derived by Shoup and Szabo. \cite{SHOUP_82}
The 4th term has been shown previously by applying a scattering method. (Eq. (29) of ref. \onlinecite{Rajendran_99})

\section{Results}
\label{sec:Results}

In the numerical calculations, Eq.~(\ref{eq:F17}) is discretized, together with
Eq.~(\ref{eq:F1}), using a standard scheme to obtain $\hat{h}(r,s)$.
The resulting $\hat{h}(r,s)$ is then substituted into Eq.~(\ref{eqF3}) to evaluate the
Laplace-domain total current through the circular region of radius $a$ at $z=0$.
The time-dependent current is obtained by numerically inverting $\hat{I}(s)$ using the
Stehfest algorithm.\cite{Stehfest1970_47,Stehfest1970_624}

Convergent values are obtained from Eqs.~(\ref{eq:Ga6}) and (\ref{eq:Ga11}) by retaining
terms up to $t^{-15/2}$ at $Dt/a^2=1$.
For $Dt/a^2>1$, fewer recursive terms are required to reach convergence.
We note, however, that including terms beyond $t^{-15/2}$ tends to increase the predicted
current; thus, the convergence is effectively asymptotic.
Nevertheless, the resulting values agree closely with the numerical calculations, as
shown in Table~\ref{tab:1}. 
The numerical results reported here are consistent with high-accuracy evaluations available in the literature; for example, the numerical coefficients in Eq.~(\ref{eq:asymnum}) are in exact agreement with those reported in Table~1 of Ref.~\onlinecite{BIENIASZ_18}. These results therefore serve as a benchmark for assessing the present analytical approximations. 
The results are summarized in Table~\ref{tab:1}.
Both the analytical and numerical values show close agreement with Eq.~(6) of Mahon and
Oldham. \cite{Mahon_05}

\begin{table}
\caption{Time dependence of the normalized current $I(t)/I_0$ for $Dt/a^2>1$}
\label{tab:1}      
\begin{tabular}{c|c|c|c|c}
\hline\noalign{\smallskip}
$Dt/a^2\, \textemdash\footnotemark[1]$ &
High-accuracy numerical \footnotemark[2]\footnotemark[3]\footnotemark[4]   & Eq. (\ref{eq:Pade3}) &
Eq.~(6) of Mahon--Oldham \cite{Mahon_05} &
Eq.~(8) of Shoup--Szabo \cite{SHOUP_82}\\
\hline
$1$ & $1.36586 (1.3659)$ & $1.36521$ & $1.36625$ & $1.374$ \\
$1.69$ & $1.27941(1.279)$ & $1.2792$ & $1.27941$ & $1.285$ \\
$2$ & $1.25641$  &$1.25627$ & $1.25641$ & $1.26148$ \\
$3$ & $1.20871$  & $1.20865$ & $1.20871$ & $1.21245$ \\
$5$ & $1.16125$  & $1.16124$ &$1.16125$ & $1.16373$ \\
$10$ & $1.1138$  & $1.1138$ & $1.1138$ & $1.11516$ \\
\noalign{\smallskip}\hline
\end{tabular}
\footnotetext[1]{Dimensionless time of Shoup--Szabo is $4Dt/a^2$. \cite{SHOUP_82}}
\footnotetext[2]{Numerical results calculated using Eq.~(\ref{eq:F17}) together with Eq.~(\ref{eq:F1}), 
providing benchmark-level accuracy for the present comparisons. 
}
\footnotetext[3]{Convergent results of Eqs.~(\ref{eq:Ga6}) and (\ref{eq:Ga11}). 
For $Dt/a^2=1$, the recursively determined current up to the $t^{-15/2}$ time dependence is used.
These results are consistent with high-accuracy numerical evaluations reported in the literature (e.g., Ref.~\onlinecite{BIENIASZ_18}).}
\footnotetext[4]{The numerical values in parentheses indicate the results of the numerical simulations reported in Ref.~\onlinecite{BRITZ_04} for $Dt/a^2=1$, \cite{BRITZ_04,BRITZ_08,Britz_16}
and
in Ref.~\onlinecite{SHOUP_82} for $Dt/a^2=1.69$.
}
\end{table}

\begin{figure}[h]
\begin{center}
\includegraphics[width=0.5\textwidth]{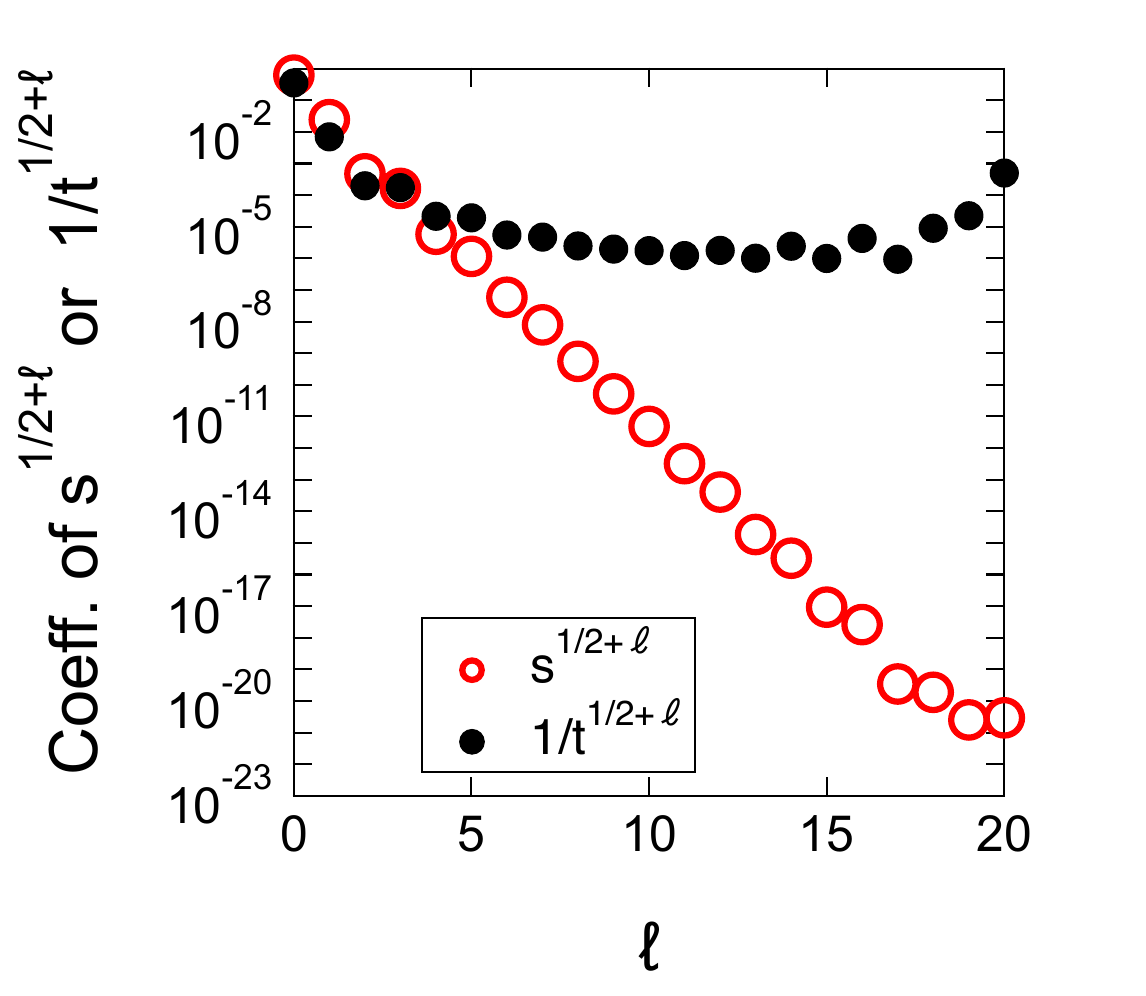}
\end{center}
\caption{
Red open circles indicate the coefficients of $s^{1/2+\ell}$ 
in the expansion of the normalized current 
$I^{(1+2\ell)}/I_{\infty} $ as a function of $\ell$ in the Laplace domain.
Black closed circles indicate the coefficients of $1/t^{1/2+\ell}$ 
in the expansion of the normalized current  
$I^{(1+2\ell)}/I_{\infty} /\Gamma(1/2-\ell)$ in the time domain.
}
\label{fig:modes}
\end{figure}

\begin{figure}[h]
\begin{center}
\includegraphics[width=0.5\textwidth]{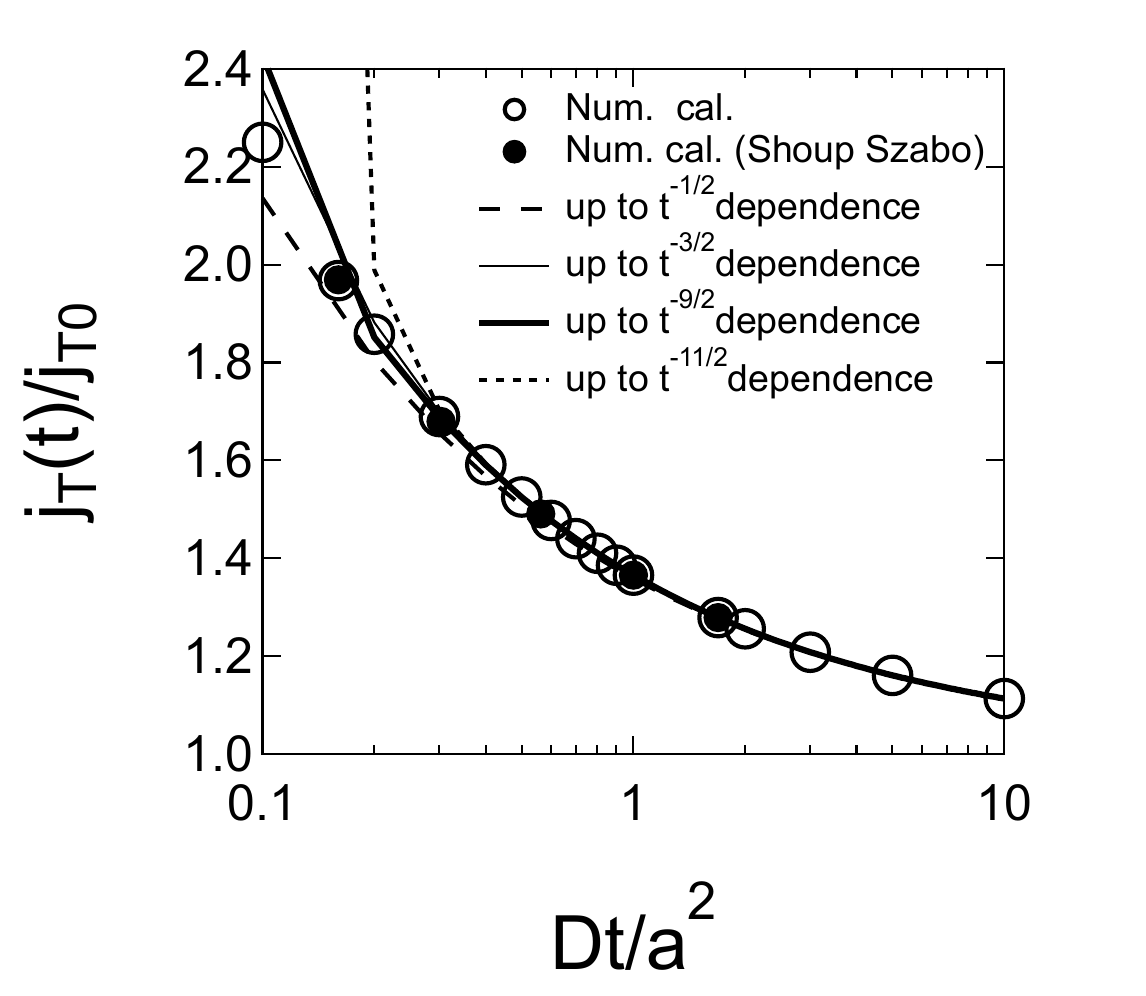}
\end{center}
\caption{
Thin long dashed, thin solid, thick solid, and thin short dashed lines indicate
$I_{\rm ip}(t)/I_{\infty} $ obtained using inverse power-law expansions up to
$1/t^{1/2}$, $1/t^{3/2}$, $1/t^{9/2}$ [Eq.~(\ref{eq:asymanal1})], and
$1/t^{11/2}$ time dependence, respectively, as given by Eq.~(\ref{eq:t13per2}).
Open circles indicate numerical results calculated by discretization of Eq.~(\ref{eq:F17}),
while closed circles indicate the numerical results of Shoup and Szabo. \cite{SHOUP_82}
}
\label{fig:modedepnd}
\end{figure}

\begin{figure}[h]
\begin{center}
\includegraphics[width=0.5\textwidth]{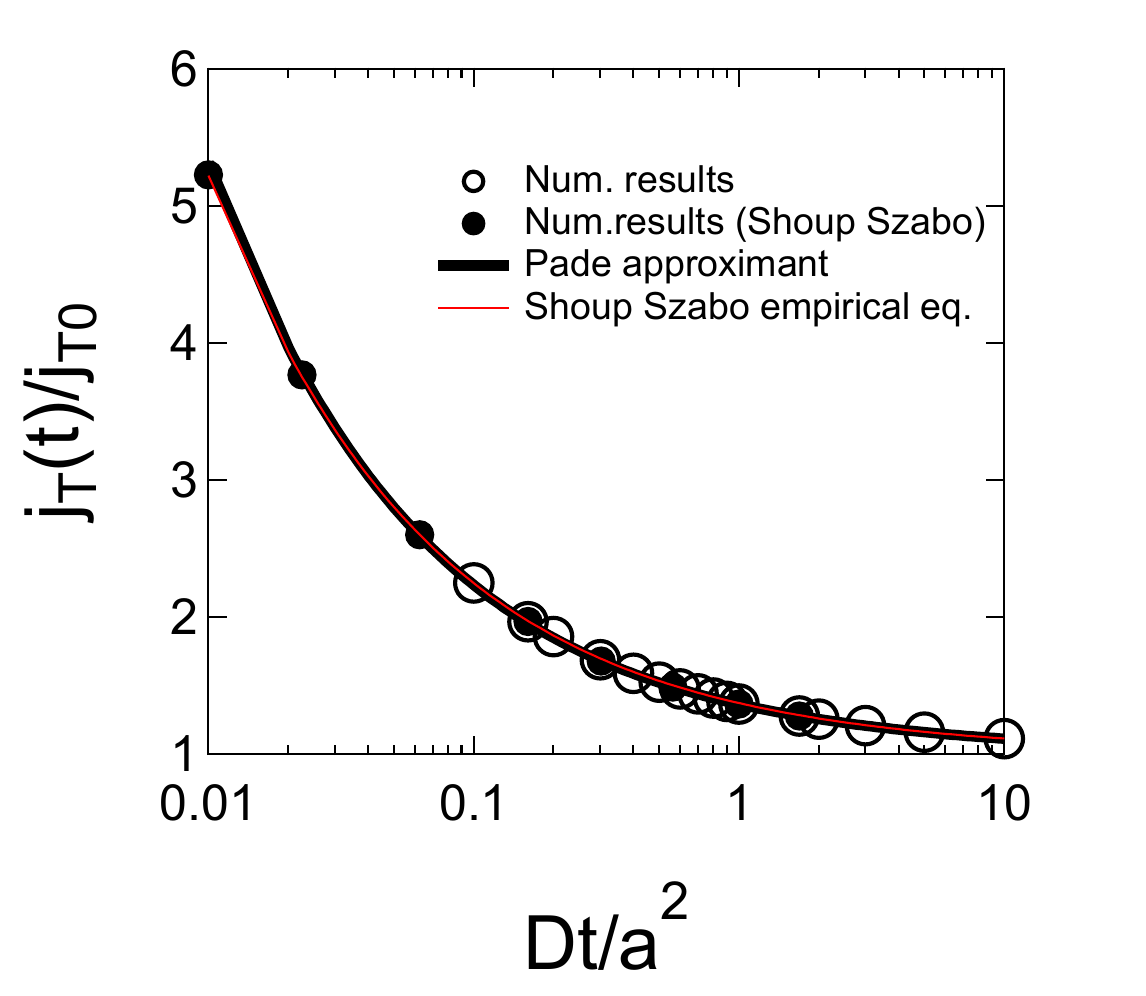}
\end{center}
\caption{
The black thick solid line shows the result of the Pad\'{e} approximant
[Eq.~(\ref{eq:Pade3})].
The red thin solid line shows the result of Shoup and Szabo
[Eq.~(\ref{eq:SZeq})]. \cite{SHOUP_82}
Open circles indicate numerical results calculated by discretization of Eq.~(\ref{eq:F17}),
and closed circles indicate the numerical results of Shoup and Szabo. \cite{SHOUP_82}
}
\label{fig:Pade}
\end{figure}

In Fig.~\ref{fig:modes}, we examine whether inverse power expansions converge
in the Laplace domain and in the time domain.
As shown in the figure,
$I^{(1+2\ell)}/I_{\infty} $ in Eq.~(\ref{eq:j_ell}), which is the coefficient of
$\left(sa^2/D\right)^{1/2+\ell}$, converges with increasing $\ell$.
In contrast,
$I^{(1+2\ell)}/I_{\infty} /\Gamma(1/2-\ell)$, which is the coefficient of
$(Dt/a^2)^{-1/2-\ell}$, decreases up to $\ell=12$ but then increases for larger $\ell$.
This difference originates from the factor $1/\Gamma(1/2-\ell)$ in the inverse
power-series expansion in the time domain.
In Table~\ref{tab:1}, we present convergent values calculated up to $\ell=7$;
these results are consistent with the numerical results obtained by discretization
of Eq.~(\ref{eq:F17}).

As shown in Fig.~\ref{fig:modedepnd}, when the plots are extended to the time region
$Dt/a^2<1$, $I(t)/I_{\infty} $ obtained using the inverse power expansion in time does
not systematically approach the numerical results with increasing $\ell$.
The best agreement in the time domain for $Dt/a^2 \gtrsim 0.2$ is obtained for
$I_{\rm ip}(t)/I_{\infty} $ given by Eq.~(\ref{eq:asymanal1}), which includes terms up to
the $1/t^{9/2}$ time dependence.

Motivated by the convergence properties of the series expansion in the Laplace domain,
we introduce a resummation method based on the Pad\'{e} approximant to obtain a
convergent analytical expression in the time domain.
First, we expand the current in the Laplace domain using
Eqs.~(\ref{eq:Ga4})--(\ref{eq:Ga6}),
\begin{align}
\frac{\hat{I}(s)}{ I_{\infty} } \approx \frac{1}{s}+\frac{2 a}{\pi  (Ds)^{1/2}}+\left(\frac{4}{\pi ^2}-\frac{1}{3}\right)\frac{a^2}{D}-
\left(\frac{1}{9}-\frac{1}{\pi^2}\right)\frac{8  a^3 s^{1/2}}{\pi  \text{D}^{3/2}} .
\label{eq:Pade1}
\end{align}
Equation~(\ref{eq:Pade1}) is obtained by performing a systematic expansion in terms of $\sqrt{s/D}\,a$, including a constant term in the Laplace domain, given by the third term on the right-hand side.
Upon inverse Laplace transformation ($1 \leftrightarrow \delta(t)$), this constant term can be neglected when considering the asymptotic time dependence.
Here, however, it is retained to interpolate the asymptotic behavior and extend the validity to short-time regimes.
We then apply a Pad\'{e} approximant of order [2,3],
\begin{align}
\frac{\hat{I}(s)}{ I_{\infty} } \approx \frac{3 \pi \left(12-\pi ^2\right)
+2 \pi^2  a\sqrt{s/D}+\pi\left(\pi ^2-8\right) a^2 s/D
}{s \left[3 \pi \left(12-\pi ^2\right)+8 \left(\pi ^2-9\right) a\sqrt{s/D} \right]} .
\label{eq:Pade2}
\end{align}
Taking the inverse Laplace transform yields the approximate analytical expression
in the time domain,
\begin{align}
\frac{I(t)}{ I_{\infty} } &\approx1+\frac{\sqrt{\pi } \left(\pi ^2-8\right)}{8 \left(\pi ^2-9\right)}
\frac{a}{ \sqrt{Dt}}- \frac{3 \left(12-\pi ^2\right)^3}{64 \left(\pi ^2-9\right)^2} 
\exp \left( c_{\rm pd}^2 \frac{Dt}{a^2}\right)
\text{erfc}\left(c_{\rm pd}\frac{\sqrt{Dt}}{a}\right),
\label{eq:Pade3}\\
&\approx 1+0.476335\frac{a}{ \sqrt{Dt}}-0.599347\exp \left( c_{\rm pd}^2 \frac{Dt}{a^2}\right)
\mathrm{erfc}\left(c_{\rm pd}\frac{\sqrt{Dt}}{a}\right),
\label{eq:Pade3n}
\end{align}
where $\mathrm{erfc}(z)$ is the complementary error function,
$\mathrm{erfc}(z)=(2/\sqrt{\pi})\int_z^\infty du\,\exp(-u^2)$, \cite{NIST} and $c_{\rm pd}$ is given by
\begin{align}
c_{\rm pd}=\frac{3 \pi  \left(12-\pi ^2\right) }{8 \left(\pi ^2-9\right)}\approx2.88616 .
\label{eq:Pade4}
\end{align}
Equation~(\ref{eq:Pade3}) can be readily evaluated using built-in implementations of special functions 
({\it e.g.}, the complementary error function) available in standard scientific software packages. 

We compare the result of Eq.~(\ref{eq:Pade3}) with the empirical equation of Shoup
and Szabo, reproduced here for completeness, \cite{SHOUP_82}
\begin{align}
\frac{I(t)}{ I_{\infty} } \approx 
\frac{\pi}{4}+\frac{a}{4}\sqrt{\frac{\pi}{Dt}}+
\left(1-\frac{\pi}{4}\right)\exp
\left[-\frac{\sqrt{\pi}/2-4 \pi^{-3/2}}{1-\pi/4}\frac{a}{2\sqrt{Dt}}\right].
\label{eq:SZeq}
\end{align}
In Fig.~\ref{fig:Pade}, very good agreement is observed among Eq.~(\ref{eq:Pade3}), the Shoup--Szabo empirical equation, and the numerical results.
Table~\ref{tab:2} summarizes numerical values of the time dependence of the current 
for $Dt/a^2<1$.
A comparison of Tables~\ref{tab:1} and \ref{tab:2} indicates that
Eq.~(\ref{eq:Pade3}) generally provides closer agreement with the numerical results than
the Shoup--Szabo empirical equation for $Dt/a^2 \gtrsim 0.2$.
For shorter times, $Dt/a^2 \lesssim 0.2$, the Shoup--Szabo empirical equation tends to show
slightly better agreement.
As shown in Appendix E, 
the typical experimentally accessible time range is $0.01 \lesssim Dt/a^2$, and  
Eq.~(\ref{eq:Pade3}) can be used to fit experimental data with accuracy comparable to that of the widely used Shoup--Szabo empirical equation [Eq.~(\ref{eq:SZeq})]. 

Equation~(\ref{eq:Pade3}) is obtained by extrapolating the asymptotic expansion to shorter
times, whereas the Shoup--Szabo empirical equation is constructed by phenomenological
interpolation between the short- and long-time limits.
Consequently, Eq.~(\ref{eq:Pade3}) is particularly suitable for describing the
intermediate- and long-time behavior, while retaining reasonable accuracy in the
short-time region.

The results reported by Mahon and Oldham are accurate; however, they are expressed in terms
of two separate equations, each applicable to a distinct time regime. \cite{Mahon_05}
Analytical expressions based on Pad\'{e} approximants applied to inverse time-series
expansions have also been proposed previously. \cite{Rajendran_99}
In such approaches, the use of higher-order Pad\'{e} approximants, such as the [5,4]
approximant, naturally leads to more involved expressions containing a larger number of
terms. \cite{Rajendran_99}
Alternatively, a simpler approximate expression has been proposed using empirical
considerations. \cite{LARAQI_11}

\begin{table}
\caption{Time dependence of the normalized current $I(t)/I_0$ for $Dt/a^2<1$}
\label{tab:2}      
\begin{tabular}{c|c|c|c|c}
\hline\noalign{\smallskip}
$Dt/a^2\, \textemdash\footnotemark[1]$ &
High-accuracy numerical \footnotemark[2] \footnotemark[3] & 
Eq. (\ref{eq:Pade3}) &
Eq.~(6) of Mahon--Oldham \cite{Mahon_05} &
Eq.~(8) of Shoup--Szabo \cite{SHOUP_82}\\
\hline
$0.1$ & $2.24999 (2.2477) $ & $2.235$ & $2.24761$ & $2.24894$ \\
$0.2$ & $1.85775 (1.8577) $ &  $1.84966$ & $1.8574$ & $1.86572$ \\
$0.3$ & $1.68964 (1.6896) $ &  $1.68456$ & $1.6892$ & $1.69948$ \\
$0.5$ & $1.5255 (1.5255) $ &  $1.52314$ & $1.52528$ & $1.53548$ \\
$0.7$ & $1.44037 (1.4404) $ &  $1.43907$ &$1.44052$ & $1.44948$ \\
$0.9$ & $1.38638 (1.3864) $ &  $1.38558$ & $1.38675$ & $1.39457$ \\
\noalign{\smallskip}\hline
\end{tabular}
\footnotetext[1]{Dimensionless time of Shoup--Szabo is $4Dt/a^2$. \cite{SHOUP_82}}
\footnotetext[2]{Numerical results calculated using Eq.~(\ref{eq:F17}) together with Eq.~(\ref{eq:F1}), providing benchmark-level accuracy for the present comparisons.
}
\footnotetext[3]{The numerical values in parentheses indicate the results of the numerical simulations reported in Ref.~\onlinecite{BRITZ_04}. \cite{BRITZ_04,BRITZ_08,Britz_16}
The present numerical results are consistent with high-accuracy numerical evaluations reported in the literature (e.g., Ref.~\onlinecite{BIENIASZ_18}).
}
\end{table}

\section{Short-time regime}
\label{sec:shorttime}

Compared with the long-time asymptotic behavior, the short-time relaxation of the current
is more difficult to analyze because an instantaneous change in interfacial ion concentration
leads to discontinuities in the current density near the edge of the disk electrode.
\cite{BIENIASZ_18}
These features originate from the mixed boundary-value nature of the problem, in which
Dirichlet boundary conditions apply on the electrode surface, whereas Neumann boundary
conditions apply on the surrounding insulating plane.
Such a configuration is characteristic of mixed boundary-value (Zaremba-type) problems.
\cite{BIENIASZ_18,zaremba_10,AKHMETGALIYEV_17}

\subsection{Mathematical analysis.}
The integral equation [Eq.~(\ref{eq:F17})] with the kernel given by Eq.~(\ref{eq:F1})
reproduces results obtained from rigorous formulations based on radial spheroidal wave
functions of the third kind. \cite{Rajendran_99,BIENIASZ_16,BIENIASZ_18}
As shown in Appendix~D, 
the numerical coefficients of the power-law expansion of the current density in Eq.~(\ref{eq:asymnum}), up to $(a/\sqrt{Dt})^{13}$, 
are in exact agreement with those reported in Table~1 of Ref.~\onlinecite{BIENIASZ_18}.
Since Eq.~(\ref{eq:F17}) with Eq.~(\ref{eq:F1}) is derived from Eq.~(\ref{eq:T5F16}), we employ
Eq.~(\ref{eq:T5F16}) to examine the short-time behavior.
This regime can be analyzed by considering the limit $s \rightarrow \infty$, in which
Eq.~(\ref{eq:T5F16}) is approximated as
\begin{align}
\frac{2}{\pi} \int_0^a dy\, \hat{h}(y,s) \int_0^\infty dk \cos(ky)
\frac{k}{\sqrt{s/D}} J_0(kr)
&\approx \frac{1}{s}
\quad \text{for } 0 \leq r < a .
\label{eq:st_1}
\end{align}

Using 10.22.59 of Ref.~\onlinecite{NIST} (1.4.13 of Ref.~\onlinecite{Duffy_08}),
\begin{align}
\int_0^\infty dk \,\sin(ky)\, J_0(kr) &=
\begin{cases}
1/\sqrt{y^2-r^2}, & y>r, \\
0, & 0\le y< r,
\end{cases}
\label{eq:sinJ0}
\end{align}
we obtain the corresponding cosine transform by differentiating with respect to $y$
(Abel-regularized), 
\begin{align}
\int_0^\infty dk \,k\cos(ky)\, J_0(kr)
&= \frac{\partial}{\partial y}
\left(\frac{1}{\sqrt{y^2-r^2}}\right)
\qquad (y>r),
\label{eq:cosJ0}
\end{align}
and the left-hand side vanishes for $0\le y<r$.
Substitution of Eq.~(\ref{eq:cosJ0}) into Eq.~(\ref{eq:st_1}) yields
\begin{align}
\lim_{\epsilon \to +0}
\frac{2}{\pi}
\int_{r+\epsilon}^a dy\, \hat{h}(y,s)
\frac{\partial}{\partial y}
\frac{1}{\sqrt{y^2-r^2}}
&\approx \frac{1}{\sqrt{sD}} .
\label{eq:st_2}
\end{align}

Performing partial integration gives
\begin{align}
\lim_{\epsilon \to +0} \frac{2}{\pi}
\left[
\left.
\hat{h}(y,s)\frac{1}{\sqrt{y^2-r^2}}
\right|_{y=r+\epsilon}^{y=a}
-\int_r^a dy\,
\frac{\partial \hat{h}(y,s)}{\partial y}
\frac{1}{\sqrt{y^2-r^2}}
\right]
&\approx \frac{1}{\sqrt{sD}} .
\label{eq:st_3}
\end{align}
Apart from the singular contribution associated with the inner endpoint $y=r$,
Eq.~(\ref{eq:st_3}) reduces to the integral term involving
$\partial \hat{h}/\partial y$.
A regular short-time solution consistent with this integral equation is given by
(see Appendix~F) 
\begin{align}
\hat{h}(y,s)=\frac{\sqrt{a^2-y^2}}{\sqrt{sD}},
\label{eq:st_5}
\end{align}
which satisfies the integral term in Eq.~(\ref{eq:st_3}) and yields the
required right-hand side $1/\sqrt{sD}$.

Substitution of Eq.~(\ref{eq:st_5}) into
$\hat{I}(s)=4DnF C_0\int_0^a dr\, \hat{h}(r,s)$ [Eq.~(\ref{eqF3})]
yields
$\hat{I}(s)=DnF C_0\pi a^2/\sqrt{sD}$.
Upon inverse Laplace transformation
($1/\sqrt{s} \leftrightarrow 1/\sqrt{\pi t}$),
this result leads to Cottrell's equation,
\cite{Cottrell_03,Soos_64,AOKI_81,SHOUP_82,AOKI_84,Phillips_90_1,OLDHAM_91,Aoki_93,KANT_94}
\[
I(t)=nF\pi a^2 C_0 \sqrt{\frac{D}{\pi t}} .
\]

This analysis demonstrates that the integral equation admits at least one admissible
short-time solution that is integrable over $0 \le r < a$ and yields a finite total current.
Since the experimentally observable quantity is the spatially integrated current, any
local modification of $\hat h(r,s)$ that does not alter this integral does not affect the
leading-order short-time behavior of $\hat I(s)$.

\subsection{Physical interpretation.}
In one-dimensional diffusion, Cottrell's equation is derived under the implicit assumption
of a uniform current density over the electrode surface.
In contrast, the present solution exhibits an explicit radial dependence of
$\hat{h}(r,s)$ and involves a kernel singularity at $y=r$ in $\hat{K}(r,y,s)$, which is
characteristic of mixed boundary-value problems.
The normal component of the flux density exhibits a radial dependence
proportional to $\sqrt{a^2-r^2}$ and vanishes at the disk edge,
reflecting the Neumann boundary condition on the surrounding insulating plane.
At the initial instant, the discontinuous change in interfacial ion concentration implies that
radial transport along the electrode surface cannot be excluded, particularly near the
disk periphery.
Nevertheless, the present analysis demonstrates that the total current follows Cottrell's 
equation, consistent with previous analytical
\cite{Soos_64,AOKI_81,SHOUP_82,AOKI_84,Phillips_90_1,OLDHAM_91,Aoki_93,KANT_94}
and numerical results. \cite{BRITZ_04}

It is worth noting that, although the formal short-time limit yields
Cottrell's equation, this regime corresponds to an idealized instantaneous
change in interfacial concentration induced by a potential
step and is often experimentally obscured by finite instrumental
response times, double-layer charging, and other non-diffusive effects.
\cite{Aoki_93,Amatore_07}
In contrast, the long-time asymptotic expansion describes the relaxation
toward the finite steady-state current $I_{\infty}\propto a$.
The excess current follows a Cottrell-type decay,
$I(t)-I_{\infty} \propto a^2\sqrt{D/t}$, with a distinct amplitude, reflecting the three-dimensional
expansion of the diffusion field required to maintain steady-state transport
to a finite disk.
This regime typically governs practical chronoamperometric measurements
at disk and microdisk electrodes and underlies widely used analytical
approximations. \cite{IKEUCHI_05}

\section{Relation to diffusional EC$'$ mechanism}
\label{sec:EC}

Owing to the initial condition $\Delta C(r,z,0)=0$, the diffusion equation in the Laplace
domain, Eq.~(\ref{eq:B1}), can be formally interpreted as a diffusion equation with a
quasi-first-order reaction, 
\begin{align}
D\nabla^2 \Delta C = K \Delta C(r,z),
\label{eq:diffK}
\end{align} 
by identifying the Laplace variable $s$ with $K$ in the steady-state limit. 
\cite{PHILLIPS_90,Rajendran_99}

A steady-state quantity can be obtained from its Laplace-domain counterpart
by multiplication by the Laplace variable $s$.
Accordingly, the steady-state current in the presence of $K$ is given by
$I_{K}=\left.
s\hat{I} (s) \right|_{s=K} $.
Using Eqs.~(\ref{eq:Pade1}) and (\ref{eq:Pade2}), we obtain
\begin{align}
\frac{I_{K}}{ I_{\infty} } &\approx 
1+\frac{2 a}{\pi } \sqrt{\frac{K}{D}}+\left(\frac{4}{\pi ^2}-\frac{1}{3}\right)\frac{a^2K}{D}-
\left(\frac{1}{9}-\frac{1}{\pi^2}\right)\frac{8 a^3 K^{3/2}}{\pi D^{3/2}}
\label{eq:diffK2_1}\\
&\approx
\frac{3 \pi \left(12-\pi ^2\right)
+2 \pi^2 a\sqrt{K/D}+\pi\left(\pi ^2-8\right) a^2 K/D
}{3 \pi \left(12-\pi ^2\right)+8 \left(\pi ^2-9\right) a\sqrt{K/D}} .
\label{eq:diffK2}
\end{align}
Equation~(\ref{eq:diffK2_1}) has been derived previously. \cite{Rajendran_99}
Results based on higher-order Pad\'{e} approximants have also been reported and shown to
provide accurate descriptions of the current. \cite{Rajendran_99,GALCERAN_99,MOLINA_16,MOLINA_19}

Equation~(\ref{eq:diffK}) is mathematically identical to the reaction--diffusion equation commonly used to describe a diffusional EC$'$ (electrochemical--catalytic or regenerative) mechanism at a disk electrode. 
In this framework, the electroactive species undergoes a homogeneous reaction characterized by a quasi-first-order rate constant $K$. 
\cite{PHILLIPS_90,Rajendran_99} 
In the classical EC$'$ framework, this reaction term gives rise to a steady-state current determined by the competition between diffusion and reaction rates. 
\cite{Nicholson_64,AMATORE_77,Bard_Faulkner}

Within the present formulation, identifying the Laplace variable $s$ with the reaction
rate constant $K$ in the steady-state limit establishes a 
formal correspondence between
the time-dependent diffusion problem and the steady-state EC$'$ description. \cite{PHILLIPS_90,Rajendran_99,Harriman_00}
Accordingly, the steady-state current $I_{K}$ obtained from
Eq.~(\ref{eq:diffK2}) 
can be interpreted as providing analytical expressions that are formally analogous to 
the EC$'$ steady-state current at a disk electrode.
The ratio $I_{K}/I_{\infty}$ therefore 
provides a measure analogous to 
the catalytic
enhancement characteristic of an EC$'$ mechanism, relative to the purely diffusion-
controlled current.

In the limiting case $K \rightarrow 0$, the reaction becomes negligible and
$I_{K}/I_{\infty} \rightarrow 1$, recovering the diffusion--controlled behavior.
Conversely, in the limit $K \rightarrow \infty$, rapid chemical regeneration leads to an
enhanced steady-state current, consistent with the catalytic EC$'$ regime discussed in
classical analyses.  \cite{Rajendran_99,MOLINA_16}
This behavior 
confirms that the present formulation reproduces the expected limiting behavior associated with diffusional EC$'$ mechanisms, 
while providing a continuous analytical description 
over intermediate values of $K$.

\section{Conclusion}
\label{sec:Conclusion}

We have investigated the relaxation of the diffusion-limited current at a disk electrode
following a change in interfacial ion concentration  induced by a potential
step over a circular reactive region
embedded in an otherwise insulating plane.
Starting from the diffusion equation with mixed Dirichlet--Neumann boundary
conditions, the problem was formulated in the Laplace domain as a modified Helmholtz
equation.
The resulting dual integral equations were reduced, via the Cooke--Sneddon method,
to a single Fredholm integral equation for an auxiliary function $\hat{h}(r,s)$.
This function directly determines the total Faradaic current, thereby providing a
transparent link between the mixed boundary-value formulation and the observable
current response.

In the steady-state limit, Saito's equation was recovered.
For the transient response, two complementary approaches were developed.
Numerical discretization of the Fredholm equation, combined with numerical Laplace
inversion, yields accurate results for $Dt/a^2 \gtrsim 0.1$.
In parallel, a systematic long-time asymptotic expansion generates inverse-power-series
corrections to the steady-state current, reproducing known results and providing
higher-order terms within a unified framework.
Because convergence is more favorable in the Laplace domain, a Pad\'{e} resummation
was introduced, leading to a simple closed-form approximation with excellent
agreement with numerical results.

The short-time analysis recovers Cottrell's equation while highlighting the role
of edge effects associated with the mixed boundary conditions.
Furthermore, identifying the Laplace variable with a quasi-first-order reaction rate
constant establishes a direct connection between the present transient formulation
and diffusional EC$'$ mechanisms.

Although the formal short-time limit yields area-proportional
Cottrell's equation, this regime corresponds to an idealized instantaneous
step and is often obscured experimentally.
In contrast, the long-time asymptotic correction follows Cottrell-type decay with a distinct amplitude 
associated with relaxation toward the finite steady-state
current of a disk electrode.
This transient reflects the three-dimensional evolution of the
diffusion field and typically governs practical chronoamperometric
measurements.

Overall, the integral-equation framework developed here provides a unified and
rigorous description of transient diffusion-limited currents at disk electrodes,
bridging steady-state and time-dependent behavior and offering reliable tools
for quantitative electrochemical analysis.

\newpage
\renewcommand{\theequation}{A.\arabic{equation}}
\setcounter{equation}{0}  
\section*{Appendix A. Derivation of Eq. (\ref{eq:T1})}
\label{appA}

In the Laplace domain, we have
\begin{align}
s \Delta\hat{C}(r,z,s)=D\nabla^2 \Delta\hat{C}, 
\label{eq:B1}
\end{align}
where $\Delta C(r,z,0)=0$ is used. 
Equation (\ref{eq:B1}) yields
\begin{align}
\nabla^2 \Delta \hat{C}(r,z,s)-\frac{s}{D} \Delta \hat{C}(r,z,s)=
\Delta_2 \Delta \hat{C}+ \frac{\partial^2}{\partial z^2} \Delta \hat{C}-\frac{s}{D} \Delta \hat{C}=0.
\label{eq:B2}
\end{align}
In the Laplace domain, the diffusion equation with zero initial conditions is the modified
Helmholtz equation, 
$\Delta_2 \Delta \hat{C}+ \frac{\partial^2}{\partial z^2} \Delta \hat{C}=\kappa^2 \Delta \hat{C}$, where $\kappa^2=s/D$. 

We introduce $\Psi (r)$ satisfying 
$\Delta_2 \Psi (r)=-k^2\Psi (r)$, 
which gives $\Psi (r)=J_0(kr)$ by noting that the Bessel equation can be written as
\begin{align}
\frac{1}{r}\frac{\partial}{\partial r} r\frac{\partial}{\partial r} J_n(kr) + k^2 J_n(kr)-\frac{n^2}{r^2} J_n(kr)=0,  
\label{eq:Bessel}
\end{align}
where $J_n(z)$ indicates the Bessel functions of the first kind, 
and the solution should be finite at $r=0$. \cite{NIST}

First, we consider the solution without taking into account the boundary conditions for the circular domain of radius $a$ at $z=0$. 
Assuming separation of variables, $\Delta \hat{C}(r,z,s)=\Psi(r)\Phi(z,s)$,
we substitute this form into Eq.~(\ref{eq:B2}) and divide by $\Psi(r)\Phi(z,s)$, which yields
\begin{align}
\frac{\Delta_2  \Psi}{\Psi}=-\frac{1}{\Phi}\frac{\partial^2}{\partial z^2}\Phi+\frac{s}{D}=-k^2 ,
\label{eq:B4}
\end{align}
and $\Phi(z,s)=\exp\left(
-|z|\sqrt{k^2+s /D} 
\right)$. 
Therefore, the solution of Eq. (\ref{eq:B1}) in the absence of boundary conditions can be written as
\begin{align}
\Delta \hat{C}(r,z,s)=J_0(kr) \exp\left(
-|z|\sqrt{k^2+s /D} 
\right) . 
\label{eq:solab}
\end{align}
It is readily verified by direct substitution that Eq.~(\ref{eq:solab}) satisfies Eq.~(\ref{eq:B1}). 
In Eq.~(\ref{eq:T1}), $A(k,s)$ is introduced to enforce the boundary conditions.

\renewcommand{\theequation}{B.\arabic{equation}} 
\setcounter{equation}{0}  
\section*{Appendix B: Derivation of Eq. (\ref{eq:F17})}
\label{appB}

Using the integral representation of the Bessel function $J_0(kr)$ \cite{NIST}
\begin{align}
J_0(kr)=\frac{2}{\pi} \int_0^r du \frac{\cos(ku)}{\sqrt{r^2-u^2}}, 
\label{eq:J0repr}
\end{align}
we rewrite Eq. (\ref{eq:T5F16}) as
\begin{align}
\int_0^r du \frac{G(u)}{\sqrt{r^2-u^2}}=\frac{1}{s}, 
\label{eq:F17_1}
\end{align}
where $G(u)$ is defined as
\begin{align}
G(u) \equiv \left( \frac{2}{\pi}\right)^2\int_0^a dy \hat{h}(y,s) \int_0^\infty d k \frac{k}{\sqrt{k^2+s /D}} \cos(ku)\cos(ky) .
\label{eq:F17_2}
\end{align}
The Abel-type integral equation in Eq. (\ref{eq:F17_1}) can be solved by a standard method. \cite{Srivastav_63,Sneddon_66,Duffy_08}
We multiply both sides of Eq. (\ref{eq:F17_1}) by 
$\int_0^x dr \, 2r/\sqrt{x^2-r^2}$. 
From the left-hand side, we obtain
\begin{align}
\int_0^x dr \frac{2r}{\sqrt{x^2-r^2}}\int_0^r du \frac{G(u)}{\sqrt{r^2-u^2}}
&= \int_0^x du \, G(u) \int_u^x dr  \frac{2r}{\sqrt{x^2-r^2}\sqrt{r^2-u^2}} \\
&= \int_0^x du \, G(u) \int_{0}^{1} dw \frac{1}{\sqrt{1-w}\sqrt{w}} \\
&=B(1/2,1/2)\int_0^x du \, G(u) =\pi \int_0^x du \, G(u),
\label{eq:Abel1}
\end{align}
where we introduced $w=(r^2-u^2)/(x^2-u^2)$; $B(p,q)$ is the Beta function, and we have used 
$B(p,q)=\Gamma(p)\Gamma(q)/\Gamma(p+q)$ with $\Gamma(1/2)=\sqrt{\pi}$. \cite{NIST}
By differentiating with respect to $x$, we obtain \cite{Duffy_08,Sneddon_66}
\begin{align}
G(x)=\frac{2}{\pi} \frac{d}{dx} 
\left[\int_0^x dr \frac{r/s}{\sqrt{x^2-r^2}}
\right] =\frac{2}{\pi s}.
\label{eq:F17_3}
\end{align}
By combining Eqs. (\ref{eq:F17_2}) and (\ref{eq:F17_3}), we obtain \cite{Duffy_08}
\begin{align}
\frac{2}{\pi}
\int_0^a dy \hat{h}(y,s) 
\int_0^\infty dk 
\frac{k}{\sqrt{k^2+s /D}} 
\cos\left(k y\right)\cos\left(k r\right)&=\frac{1}{s}  \mbox{ for } 0 \leq r < a. 
\label{eq:T10}
\end{align}
$\hat{h}(y,s)$ should be a solution of Eq. (\ref{eq:T10}) in order to satisfy the boundary condition in Eq. (\ref{eq:T5}), 
originating from Eq. (\ref{eq:T15}). 

Below, we rewrite Eq. (\ref{eq:T10}) into a form suitable for obtaining a numerical solution. 
Using (1.17.12.1 of Ref.~\onlinecite{DLMF} or 2.1.18 of Ref.~\onlinecite{Duffy_08}),
\begin{align}
\delta(r-y)=\frac{2}{\pi}\int_0^\infty dk\,\cos(kr)\cos(ky),
\label{eq:T7}
\end{align}
we obtain
\begin{align}
\int_0^a dy\, h(y)\,\frac{2}{\pi}\int_0^\infty dk\,\cos(ky)\cos(kr)
&=\int_0^a dy\, h(y)\,\delta(r-y)
\nonumber\\
&=
\begin{cases}
h(r), & 0\le r<a,\\
0, & r>a. 
\end{cases}
\label{eq:T8}
\end{align}
By combining Eqs.~(\ref{eq:T10}) and (\ref{eq:T8}), we obtain Eq.~(\ref{eq:F17}).

\newpage
\renewcommand{\theequation}{C.\arabic{equation}} 
\setcounter{equation}{0}  
\section*{Appendix C: Dimensionless formulation}
\label{appC}

For generality, we reformulate the problem using the dimensionless variables $x=r/a$ and $\xi=ka$.
Equation (\ref{eq:T1}) can be expressed using $z_{\rm n}=z/a$ as,
\begin{align}
\Delta \hat{C}(r,z,s) =
\int_0^\infty d\xi \, \bar{A}(\xi,s) J_0(\xi x)
\exp\!\left(-z_{\rm n}\sqrt{\xi^2 + s a^2/D}\right). 
\label{eq:T1_dl}
\end{align}

Similarly, the equations corresponding to Eqs.~(\ref{eq:T4})--(\ref{eq:T6}) are expressed as, 
\begin{align}
\bar{g}(\xi,s) &= \sqrt{\xi^2+s a^2/D}\,\bar{A}(\xi,s)/C_0 ,
\label{eq:T4_dl}\\
\int_0^\infty d \xi L(\xi,s)\bar{g}(\xi,s) \, J_\nu(\xi x) &=x^{\nu} \mbox{ for } 0 \leq x < 1 ,
\label{eq:gm1}\\
\int_0^\infty d \xi \, \bar{g}(\xi,s) J_\nu(\xi x ) &=0 \mbox{ for } 1 < x < \infty ,
\label{eq:gm2}
\end{align}
where the Laplace transform of a function expressed in dimensionless variables ({\it e.g.}, $f$) is denoted by $\bar{f}$.
In our case, $L(\xi,s)\equiv s/\sqrt{\xi^2+s a^2/D}$ and $\nu=0$. 

According to Cooke (1956) (4.3.24--4.3.27, p.~213 of Ref.~\onlinecite{Duffy_08}), \cite{COOKE_56}
the solution of Eqs.~(\ref{eq:gm1}) and (\ref{eq:gm2}) is given by
\begin{align}
\bar{g}(\xi,s) &= \frac{2^\beta \Gamma(\nu+1)}{\Gamma(\nu-\beta+1)} \xi^{1+\beta} \int_0^1 dt \, \bar{f}(t,s) t^{\alpha+1} J_{\nu-\beta}(\xi t) .
\label{eq:gm3}
\end{align}
where $\Gamma(z)$ is the Gamma function; \cite{NIST}
$\bar{f}(x,s)$ satisfies
\begin{align}
\bar{f}(x,s)+x^{-\alpha} \int_0^1 dt\, t^{\alpha+1} \bar{f}(t,s)
\left[\int_0^\infty d \xi \, 
\left(a \xi^{2\beta} L(\xi,s)-1
\right) \xi J_{\nu -\beta}(\xi t) J_{\nu -\beta}(\xi x)
\right] =a x^{\nu-\alpha-\beta},
\label{eq:gm4}
\end{align}
where $a$, $\alpha$, and $\beta$ are arbitrary parameters, provided that $0<\beta<1$ and $-1<\nu-\beta$ for real $\beta$ and $\nu$. 

We choose $\beta=1/2$, so that
$J_{-1/2}(\xi t)=\sqrt{2/(\pi \xi t)} \cos(\xi t)$. 
The other parameters are chosen as $a=1/s$ and $\alpha=-1/2$. 

Accordingly, we introduce $\bar{h}(x,s)$ satisfying 
\begin{align}
\bar{g}(\xi,s)=\frac{2}{\pi} \xi \int_0^1 dx \, \bar{h}(x,s) \cos(\xi x) ,
\label{eq:gm5}
\end{align} 
and
\begin{align}
\bar{h}(x,s)+ 
\frac{2}{\pi}
\int_0^1 dy \, \bar{h}(y,s) 
\int_0^\infty d \xi 
\left[\frac{\xi}{\sqrt{\xi^2+s a^2/D}}-1 
\right]
\cos(\xi y)\cos(\xi x) =\frac{1}{s}
\mbox{ for } 0 \leq x < 1. 
\label{eq:gm6}
\end{align}

Using the dimensionless formulation, Eqs.~(\ref{eqF2}) and (\ref{eqF3}) become
\begin{align}
\hat{I} (s)&=2\pi D a nF C_0  \int_0^\infty d \xi \bar{g}(\xi,s) \frac{1}{\xi} J_1 (\xi) ,
\label{eqF2_dl}\\
&=4D nF C_0 \int_0^1 dx \, \bar{h}(x,s).
\label{eqF3_dl}
\end{align}

\renewcommand{\theequation}{D.\arabic{equation}} 
\setcounter{equation}{0}  
\section*{Appendix D: Fast evaluation of the iterated functions and the total current}
\label{appD}

The recursive relation for the iterated functions,
\begin{align}
h^{(\ell)} (r) &= \sum_{n=1}^\ell \int_0^a dy \, \bar{K}^{(n)} (r,y)\, h^{(\ell-n)} (y),
\qquad \ell\geq 1,
\label{eq:Ga6_apB}
\end{align}
with $h^{(0)}(r)=1$ and $\bar{K}^{(n)}=K^{(n)}/a^n$,
is identical to Eq.~(\ref{eq:Ga6}) in the main text.
A direct symbolic implementation of this recursion is computationally expensive,
since each iteration involves repeated symbolic integrations.
In this Appendix, we present an efficient computational scheme that exploits the
polynomial structure of the iterated functions and avoids symbolic integration
altogether.

Since $K^{(n)} (r,y)$, given by Eq.~(\ref{eq:Ga4}), is a polynomial in $r$ with
maximum degree $n-1$, it follows that each iterated function $h^{(\ell)}(r)$ is also
a polynomial in $r$ with maximum degree $\ell$, given that $h^{(0)}(r)=1$.
Accordingly, we may write
\[
h^{(\ell)}(r)=\sum_{m=0}^\ell c_m\,r^{m}.
\]
Furthermore, the kernel $\bar{K}^{(n)}(r,y)$ is a linear combination of the
polynomial factors $(r+y)^{n-1}$, $(r-y)^{n-1}$, and $|r-y|^{n-1}$
[cf.\ Eq.~(\ref{eq:Ga4})].
By expanding these factors binomially, the integrand in
Eq.~(\ref{eq:Ga6_apB}) is reduced to a finite sum of monomials in $y$.

As a consequence, all integrals appearing in the recursion can be expressed in
terms of the elementary moments
\begin{align}
M_a(m)&=\int_0^a y^{m}\,dy=\frac{a^{m+1}}{m+1},\\
M_r(m)&=\int_0^r y^{m}\,dy=\frac{r^{m+1}}{m+1},\\
M_{r\to a}(m)&=\int_r^a y^{m}\,dy=\frac{a^{m+1}-r^{m+1}}{m+1}.
\end{align}
The absolute-value contribution is treated exactly by splitting the integration
domain at $y=r$,
\[
\int_0^a |r-y|^{n-1}y^{m}\,dy
=
\int_0^r (r-y)^{n-1}y^{m}\,dy
+\int_r^a (y-r)^{n-1}y^{m}\,dy,
\]
after which each term is again reduced to the precomputed moments $M_r$ and
$M_{r\to a}$ following binomial expansion.

In practice, the evaluation of the right-hand side of
Eq.~(\ref{eq:Ga6_apB}) for a fixed $n$ proceeds as follows:
(i) expand $h^{(\ell-n)}(y)$ into monomials;
(ii) expand the kernel factors $(r\pm y)^{n-1}$ and $(y-r)^{n-1}$ binomially;
and
(iii) replace all resulting integrals of $y^{m}$ over the relevant interval by
$M_a(m)$, $M_r(m)$, or $M_{r\to a}(m)$.
This procedure yields an explicit polynomial expression for $h^{(\ell)}(r)$ without
invoking symbolic integration.
Iterating this procedure efficiently implements the recursion
(\ref{eq:Ga6_apB}) and enables the computation of high-order iterated functions.

Once $h^{(\ell)}(r)$ has been obtained in polynomial form, the coefficient entering
the asymptotic expansion of the normalized current is evaluated as
\[
\frac{I^{(\ell)}}{I_{\infty}}=\frac{1}{a}\int_0^a h^{(\ell)}(r)\,dr,
\]
where $I_{\infty} =4DanFC_0$.
This integral is again computed using the monomial rule
$\int_0^a r^m\,dr=a^{m+1}/(m+1)$.
The truncated long-time expansion of the normalized current is then
constructed using,
\[
\frac{I(t)}{I_{\infty}}\approx
1+\sum_{\ell=0}^{\ell_{\max}}
\frac{I^{(1+2\ell)}}{I_{\infty} }
\,
\frac{(Dt/a^2)^{-1/2-\ell}}{\Gamma\!\left(\tfrac12-\ell\right)} ,
\label{eq:eq:ApBJt}
\]
while the corresponding expression in the Laplace domain is
\begin{align}
\frac{\hat{I}(s)}{I_{\infty}}\approx
\frac{1}{s}\left(1+\sum_{\ell=0}^{\ell_{\max}}
\frac{I^{(1+2\ell)}}{I_{\infty} }
\,
\left(sa^2/D\right)^{1/2+\ell} \right). 
\label{eq:ApBJs}
\end{align}

As an illustration of the efficiency of this approach, we report the systematic asymptotic expansion up to a certain order in the Laplace domain,
\begin{multline}
\frac{\hat{I}(s)}{ I_{\infty}} \approx \frac{1}{s}+\frac{2 a}{\pi  (Ds)^{1/2}}-
\left(\frac{1}{9}-\frac{1}{\pi^2}\right)\frac{8  a^3 s^{1/2}}{\pi  \text{D}^{3/2}}
+\left(\frac{142}{675}
-\frac{16}{3 \pi^2}+\frac{32}{\pi^4}
\right)
\frac{a^5 s^{3/2}}{\pi  D^{5/2}}+\\
\left(-\frac{1136}{33075}+\frac{3872}{2025 \pi^2}-\frac{256}{9 \pi^4}+\frac{128}{\pi^6}
\right)
\frac{a^7 s^{5/2}}{\pi  D^{7/2}}+
\left(
\frac{19046}{4465125}-\frac{86336}{178605\pi^2}+\frac{608}{45\pi^4}-\frac{1280}{9\pi^6}+\frac{512}{\pi^8}
\right)
\frac{a^{9} s^{7/2}}{\pi D^{9/2}}+\\
\left(
-\frac{658136}{1620840375 }+\frac{1280408}{13395375 \pi^2}-\frac{161408}{35721 \pi^4}+\frac{56576}{675 \pi^6}-
\frac{2048}{3 \pi^8}+\frac{2048}{\pi^{10}}
\right)
\frac{a^{11} s^{9/2}}{\pi  D^{11/2}} +\\
\left(\frac{57122836 }{1917454163625}
-\frac{400816}{25727625 \pi ^2}
+\frac{34023424}{28704375 \pi ^4}
-\frac{4417024}{127575 \pi ^6}
\right. \\ \left.
+\frac{971264}{2025 \pi ^8}
-\frac{28672}{9 \pi ^{10}}
+\frac{8192}{\pi ^{12}}
\right)
\frac{a^{13} s^{11/2}}{\pi  D^{13/2}} .
\label{eq:t13per2s}
\end{multline}
Applying the inverse Laplace transformation via the Tauberian theorem yields the
corresponding expression in the time domain,
\begin{multline}
\frac{I(t)}{ I_{\infty} } \approx 1+ \frac{2a}{\pi^{3/2} \left( Dt \right)^{1/2}}
+\left(\frac{1}{9}-\frac{1}{\pi^2} \right) \frac{4a^3}{\left(\pi D t\right)^{3/2}} +
\left(\frac{71}{450 }-\frac{4}{\pi ^{2}}+\frac{24}{\pi ^{4}}\right) \frac{a^5}{\pi^{3/2}(Dt)^{5/2}}+
\\
\left(\frac{71}{2205}-\frac{242}{135 \pi ^{2}}+\frac{80}{3 \pi ^{4}}-\frac{120}{\pi ^{6}}\right)
\frac{2a^7}{\pi^{3/2}(Dt)^{7/2}}
+
\left(\frac{9523}{340200}-\frac{5396}{1701 \pi ^{2}}+\frac{266}{3 \pi ^{4}}-\frac{2800}{3 \pi ^{6}}+\frac{3360}{\pi ^{8}}
\right)\frac{a^9}{\pi^{3/2}(Dt)^{9/2}}
+\\
\left(\frac{82267}{6860700}-\frac{160051}{56700 \pi ^{2}}+\frac{25220}{189 \pi ^{4}}-\frac{12376}{5 \pi ^{6}}+\frac{20160}{\pi ^{8}}-\frac{60480}{\pi ^{10}}
\right)\frac{a^{11}}{\pi^{3/2}(Dt)^{11/2}} +\\
\left(\frac{14280709}{2951348400}
-\frac{25051}{9900 \pi ^{2}}
+\frac{5847776}{30375 \pi ^{4}}
-\frac{759176}{135 \pi ^{6}}
+\frac{1168552}{15 \pi ^{8}}
-\frac{517440}{\pi ^{10}}
+\frac{1330560}{\pi ^{12}}
\right)
\frac{a^{13}}{\pi^{3/2}(Dt)^{13/2}} .
\label{eq:t13per2}
\end{multline}

In order to compare Eq.~(\ref{eq:t13per2}) with previous results,
we express Eq.~(\ref{eq:t13per2}) by truncating the irrational numbers originating from
$\pi$ as
\begin{multline}
\frac{I(t)}{ I_{\infty} } \approx 
1+0.359174\sqrt{\frac{a^2}{Dt}}+0.00703258\left(\frac{a^2}{Dt}\right)^{3/2}
-0.000201745 \left(\frac{a^2}{Dt}\right)^{5/2}-0.000175244\left(\frac{a^2}{Dt}\right)^{7/2}+
\\
0.0000217006\left(\frac{a^2}{Dt}\right)^{9/2}+
0.0000195802\left(\frac{a^2}{Dt}\right)^{11/2}-
5.50927 \times 10^{-6}\frac{a^{13}}{(Dt)^{13/2}} . 
\label{eq:asymnum}
\end{multline}
The numerical values are identical to those reported in Table~1 of Ref.~\onlinecite{BIENIASZ_18}.
In Ref.~\onlinecite{BIENIASZ_18}, these values are obtained by evaluating the exact expression
for the current expanded in terms of radial spheroidal wave functions of the third
kind. \cite{Rajendran_99,BIENIASZ_16,BIENIASZ_18}
The agreement between our results and those reported in Ref.~\onlinecite{BIENIASZ_18}
indicates the correctness of our calculations.
The terms from $1/t^{5/2}$ up to the $1/t^{9/2}$ time dependence have also been reported
numerically. \cite{AOKI_84,Mahon_05}
Our results are consistent, up to the first three digits of the coefficient of $1/t^{5/2}$,
with Eq.~(6) of Mahon and Oldham~\cite{Mahon_05}, after multiplying the coefficient by
$4/\pi$ to account for the difference in the normalization of the current; 
according to the normalization reported in Ref.~\onlinecite{Mahon_05}, 
the numerical coefficients are $4/\pi$, $8/\pi^{5/2}$, 
$0.00895416$, $0.00025687$, $0.000223127$, $0.0000276301$, and $0.0000249303$.

\renewcommand{\theequation}{E.\arabic{equation}} 
\setcounter{equation}{0}  
\section*{Appendix E: Experimental verification of the transient current prediction of Eq.~(\ref{eq:Pade3})}
\label{appE}

Here, we present an experimental verification of the transient current predictions using reported potential step chronoamperometry data shown in Fig.~6(a) of Ref.~\onlinecite{IKEUCHI_00}. 
We analyze dpotential step chronoamperometry data of a typical reversible redox couple 
([Fe(CN)$_6$]$^{4-}$/[Fe(CN)$_6$]$^{3-}$) (1 mol m$^{-3}$) in aqueous KCl solution (1 mol dm$^{-3}$) at a Pt disk electrode (radius 0.101 mm) at \qty{25}{\degreeCelsius}. 
The other experimental conditions are given in Ref.~\onlinecite{IKEUCHI_00}. 

\begin{figure}[h]
\begin{center}
\includegraphics[width=0.4\textwidth]{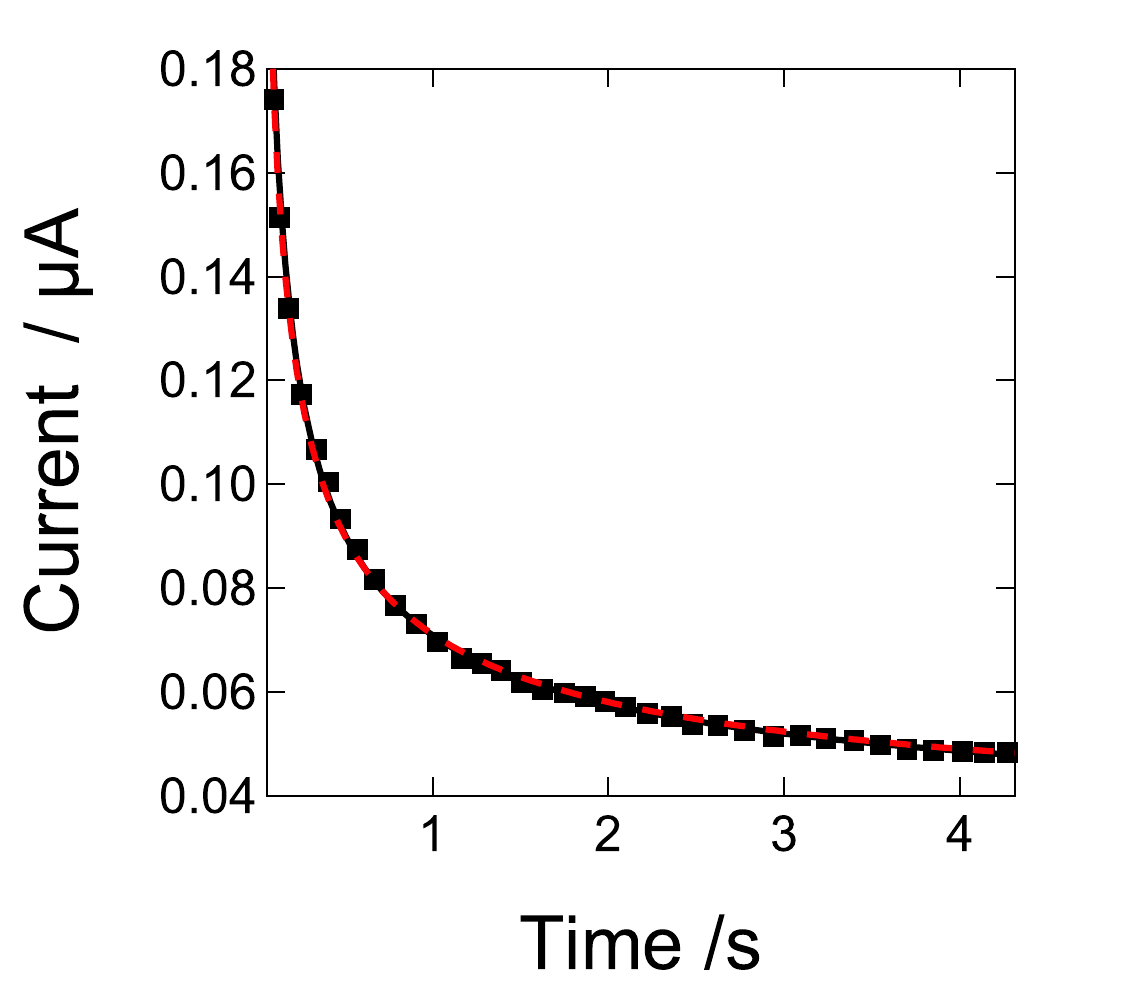}
\end{center}
\caption{
The dots represent the absolute value of the cathodic current transient for 
potential step chronoamperometry of [Fe(CN)$_6$]$^{3-}$ 
(1 mol m$^{-3}$) in aqueous KCl solution (1 mol dm$^{-3}$) at a Pt disk electrode 
(radius 0.101 mm) at \qty{25}{\degreeCelsius}, reproduced from Ref.~\onlinecite{IKEUCHI_00} (Fig.~6(a)). 
The black thick solid line shows the fit to the Pad\'e approximant [Eq.~(\ref{eq:Pade3})], 
and the red thick dashed line shows the fit to the Shoup--Szabo equation [Eq.~(\ref{eq:SZeq})]~\cite{SHOUP_82}. 
Reproduced from Ref.~\onlinecite{IKEUCHI_00} with permission from Elsevier. 
}
\label{fig:Ikeuchi}
\end{figure}

The diffusion coefficients and the transient currents analyzed using Eq.~(\ref{eq:Pade3}) are compared with those estimated using the Shoup--Szabo equation [Eq.~(\ref{eq:SZeq})]. 
The experimentally accessible time range corresponds to $0.01 \lesssim Dt/a^2 \lesssim 0.2$, 
in which Eq.~(\ref{eq:Pade3}) is expected to provide sufficient accuracy, as demonstrated in Fig.~\ref{fig:Pade}. 
Note that Eq.~(\ref{eq:Pade3}) is constructed using a Pad\'e approximant that refines the long-time asymptotic decay, 
whereas the Shoup--Szabo equation interpolates between Cottrell's equation (valid in the short-time regime) and the long-time asymptotic decay. 
In the present time range, the short-time regime described by Cottrell's equation is not dominant (it typically appears for $Dt/a^2 \lesssim 0.01$), 
making the Pad\'e-based expression conceptually more suitable.

In both Eqs.~(\ref{eq:Pade3}) and (\ref{eq:SZeq}), we introduce a single fitting parameter $C_{\rm F}=D/a^2$ and 
express $I_{\infty}=4D a nF C_0$ as 
$I_{\infty}=4C_{\rm F} a^3 nF C_0$. 

Both models were fitted to the data in the time window $0.20 \le t \le 4.00~\mathrm{s}$, 
excluding the unresolved early-time transient region.

The diffusion coefficients obtained from the two models are:
\begin{align}
D &= (7.35 \pm 0.02)\times10^{-10}\ \mathrm{m^2\,s^{-1}}, \\
D &= (7.33 \pm 0.03)\times10^{-10}\ \mathrm{m^2\,s^{-1}},
\end{align}
using Eq.~(\ref{eq:Pade3}) and the Shoup--Szabo equation, respectively, 
where the uncertainty was obtained from the standard deviation of the residuals within the fitting window. 

Accordingly, both models reproduce the experimental data with comparable accuracy in the measured time range, as shown in Fig.~\ref{fig:Ikeuchi}.   
The remaining deviations are attributed to digitization of the original figure and the unresolved early-time transient. 

For comparison, the diffusion coefficient reported in the original experimental work is 
$7.74\times10^{-10}\ \mathrm{m^2\,s^{-1}}$, averaged over four measurements~\cite{IKEUCHI_00}. 
This value is in good agreement with those obtained here, confirming that both models provide reliable estimates of the diffusion coefficient within the experimentally accessible time range.

\renewcommand{\theequation}{F.\arabic{equation}} 
\setcounter{equation}{0}  
\section*{Appendix F: Verification of Eq.~(\ref{eq:st_5})}
\label{appF}

To verify that Eq.~(\ref{eq:st_5}) satisfies the regular (integral) term in
Eq.~(\ref{eq:st_3}), we note that
\begin{align}
\frac{\partial \hat{h}(y,s)}{\partial y}
=-\frac{1}{\sqrt{sD}}\frac{y}{\sqrt{a^2-y^2}} .
\label{eq:E1}
\end{align}
Therefore,
\begin{align}
-\int_r^a dy\,
\frac{\partial \hat{h}(y,s)}{\partial y}\,
\frac{1}{\sqrt{y^2-r^2}}
=
\frac{1}{\sqrt{sD}}
\int_r^a dy\,
\frac{y}{\sqrt{a^2-y^2}}\,
\frac{1}{\sqrt{y^2-r^2}} .
\label{eq:E2}
\end{align}

Using the integral formula
\begin{align}
\int_r^a dy\, \frac{y}{\sqrt{a^2-y^2}} \frac{1}{\sqrt{y^2-r^2}}=\frac{\pi}{2},
\label{eq:st_4a}
\end{align}
obtained by the change of variables $t=y^2$ and the Beta-function identity 
(Eq.~(3.196.3) of Ref.~\onlinecite{gradshteyn_96})
\cite{NIST}
\begin{align}
\int_\alpha^\beta dt\, (\beta-t)^{\nu-1}(t-\alpha)^{\mu-1}
=(\beta-\alpha)^{\mu+\nu-1} B(\mu,\nu),
\label{eq:st_4b}
\end{align}
with $\mu=\nu=1/2$ and $B(1/2,1/2)=\Gamma(1/2)^2=\pi$,
Eq.~(\ref{eq:E2}) becomes
\[
-\int_r^a dy\,
\frac{\partial \hat{h}(y,s)}{\partial y}\,
\frac{1}{\sqrt{y^2-r^2}}
=
\frac{\pi}{2}\frac{1}{\sqrt{sD}}.
\]
Substituting this result into Eq.~(\ref{eq:st_3}) yields
$1/\sqrt{sD}$ for the regular contribution,
which is consistent with Eq.~(\ref{eq:st_5}).

\end{document}